\documentclass[%
 aip,
 amsmath,amssymb,
 reprint,%
]{revtex4-1}

\usepackage{graphicx}
\usepackage{dcolumn}
\usepackage{bm}

\usepackage[utf8]{inputenc}
\usepackage[T1]{fontenc}
\usepackage{mathptmx}
\usepackage{etoolbox}

\makeatletter
\def\@email#1#2{%
 \endgroup
 \patchcmd{\titleblock@produce}
  {\frontmatter@RRAPformat}
  {\frontmatter@RRAPformat{\produce@RRAP{*#1\href{mailto:#2}{#2}}}\frontmatter@RRAPformat}
  {}{}
}%
\makeatother
\begin{document}

\preprint{AIP/123-QED}

\title[Effective Electronic Structure]{Effective Electronic Structure of Monoclinic $\beta-(Al_xGa_{1-x})_2O_3$ alloy semiconductor}
\author{Ankit Sharma}
 \altaffiliation[Also at ]{Department of Electrical Engineering, University at Buffalo}
\author{Uttam Singisetti}%
 \email{ankitsha@buffalo.edu}
\affiliation{ 
University at Buffalo, State University of New York
}%


\date{\today}

\begin{abstract}
In this article, the electronic band structure $\beta-(Al_xGa_{1-x})_2O_3$ alloy system is calculated with $\beta-Ga_2O_3$ as the bulk crystal. The technique of band unfolding is implemented to obtain the effective bandstructure \textit{(EBS)} for aluminium fractions varying between 12.5\% and 62.5\% with respect to the gallium atoms. A 160 atom supercell is used to model the disordered system that is generated using the technique of special quasirandom structures which mimics the site correlation of a truly random alloy and reduces the configurational space that arises due to the vast enumeration of alloy occupation sites. The impact of the disorder is then evaluated on the electron effective mass and bandgap which is calculated under the generalized gradient approximation \textit{(GGA)}. The EBS of disordered systems gives an insight into the effect of the loss of translational symmetry on the band topology which manifests as band broadening and can be used to evaluate disorder induced scattering rates and electron lifetimes. This technique of band unfolding can be further extended to alloy phonon dispersion and subsequently phonon lifetimes can also be evaluated from the band broadening.
\end{abstract}

\maketitle

\section{Introduction}
Gallium oxide $(Ga_2O_3)$ is a promising wide bandgap semiconductor material system with potential applications ranging from power and RF devices to deep UV photodetectors. The monoclinic $\beta$-phase is the most thermodynamically stable phase among all the phases\cite{poly1,thermstb1,thermstb2}. $\beta-Ga_2O_3$ has a large bandgap $(4.6-4.9eV)$\cite{gap1,gap2,thermstb}, a high critical breakdown field strength $(8MV/cm)$\cite{breakdown1} and high electron saturation velocity\cite{satvelocity1} resulting in higher Baliga's figure of merit $(BFoM)$\cite{bfom1,bfom2} and Johnson's figure of merit $(JFoM)$ compared to GaN and SiC, making it attractive for power and RF devices. With maturity in material growth and device fabrication techniques for $\beta-Ga_2O_3$, devices such as lateral FET's with $8kV$\cite{brkdwn1} breakdown and MESFET's\cite{mesfet1,mesfet2,mesfet3,mesfet4} with high $f_T$ have been successfully fabricated. Recently, $p$-type doping has also been achieved using magnesium $(Mg)$ or nitrogen $(N)$ as dopants which act as deep acceptors\cite{ptype1} using the technique of ion implantation. Unlike \textit{n}-type, \textit{p}-type doping is difficult in $\beta-Ga_2O_3$ material system due to the flat valence bands contributed mainly by oxygen \textit{2p} orbitals. This further broadens the scope of its potential applications.

$\beta-Ga_2O_3$ can be alloyed with aluminium (\textit{Al}) which increases the bandgap with increasing aluminium fraction. The large bandgap $(Al_xGa_{1-x})_2O_3$ alloy can be used to grow of $AlGaO/GaO$ heterostructures and design devices\cite{algao5} to utilize the benefits of high electron mobility of the 2-D electron gas formed at the heterojunction. The theoretically calculated low field 2DEG electron mobility in $AlGaO/GaO$ heterostrucutres is around $500cm^2/Vs$\cite{2deg1} which is higher than the bulk mobility of $200cm^2/Vs$\cite{lfmob1,lfmob2} at room temperature which is primarily limited by the polar optical phonon scattering. As a result of successful growth of $AlGaO/GaO$ heterostructures, HFET's with high transconductance and cutoff frequency have also been demonstrated\cite{hfet1}. With the recent reports indicating the maturity in growth of $AlGaO$\cite{algao1,algao2,algao3,algao4,algao5} alloys and their corresponding theoretical studies, it has become ever more pertinent to develop and understand the electronic dispersion and transport property of this alloy from first principles. In this article, the effective electronic dispersion of the $AlGaO$ alloy system is calculated from which properties such as electron effective mass, variation of band gap is obtained which will be instrumental in extending the study to develop a low field electron transport model in disordered system. 

Traditionally, the theoretical calculations for alloys were carried under approximations such as the VCA\cite{vca} (Virtual crystal approximation) where the atomic sites are substituted with hybrid atoms with mixed pesudopotentials that are generated by combining the elements in the exact stoichiometric ratio as in the parent material. Although this approach is computationally efficient as it reduces the system size significantly, it fails to capture the effect of local disorder as it imposes artifical translational symmetry to the system. Another approximation frequently used for the electronic dispersion calculation is the CPA\cite{cpa} (Coherent potential approximation) which again has the limitation of imposing order in the random alloy\cite{cpa1,cpa2}. The well established method of electronic dispersion calculation using density functional theory (DFT) and planewave expansion in case of alloy is challenging due to the loss of translational symmetry imposed by local disorder. This work focuses on the calculation of the electronic structure of $AlGaO$ alloy system from first principles using supercells and the technique of \textit{band-unfolding} developed by Zunger \textit{et.al.}\cite{bandunfold1,bandunfold2}. This majority representation technique was successfully employed for $InGaN$ and $GaNP$ alloy systems\cite{bandunfold1,bandunfold2} to generate the effective bandstructure $(EBS)$.

\begin{figure*}[h]
    \includegraphics[width=0.85\linewidth]{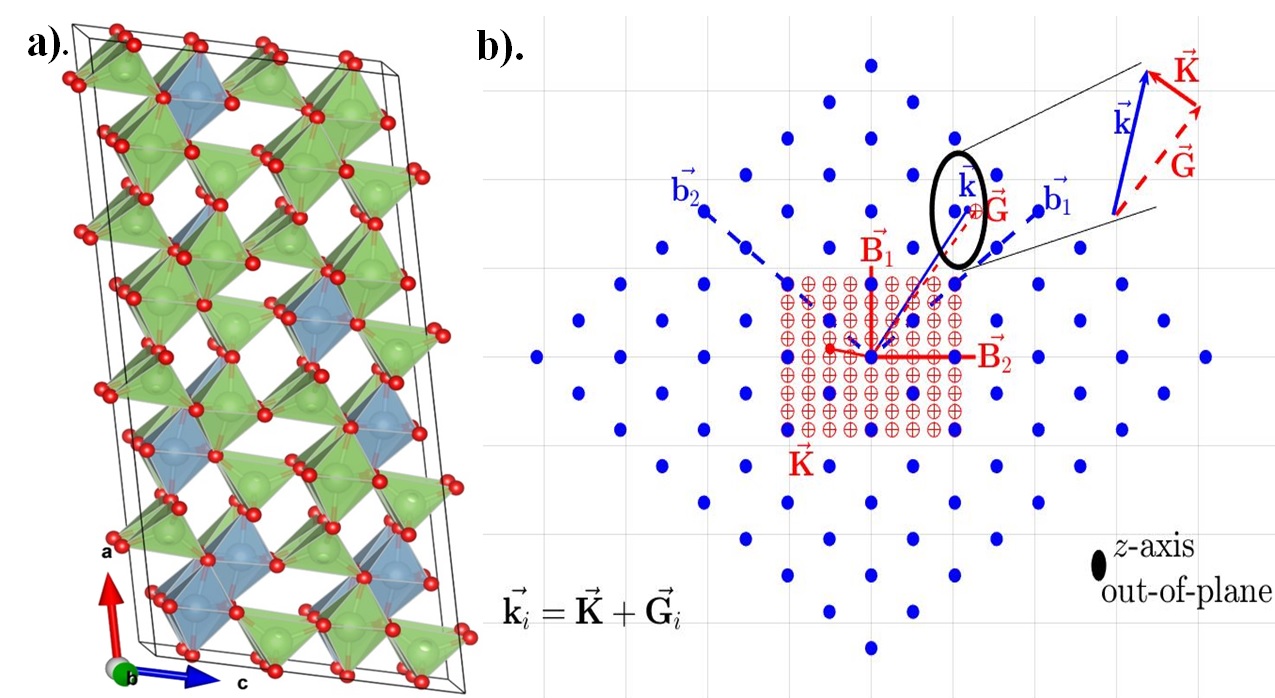}
    \caption{\label{fig:structure} a). The 160 atom $AlGaO$ supercell rendered using VESTA\cite{vesta} with $18.75\%$ $Al$ content represented in \textit{blue}, $Ga$ in \textit{green} and $O$ in \textit{red}. b). The vector representation of eqn.\ref{eq:1}where the red markers represent the $SC$ reciprocal lattice, the blue markers for $PC$ reciprocal lattice. The vectors hold their usual meaning as described in eqn.\ref{eq:1} }
\end{figure*}

In order to overcome the computational challenge due to the loss of translational symmetery with aluminium incorporation, the alloy disorder was modeled using supercell with gallium oxide conventional cell as the fundamental repeating unit. $\beta-Ga_2O_3$ has a monoclinic crystal structure with a 10 atom basis and has $C2/m$ space group symmetry. The primitive unit cell contains 2 formula units of $Ga_2O_3$ with 2 pairs of $Ga$ atoms having tetrahedral and the other two with octahedral co-ordinations respectively. The supercell used in this work is generated from the conventional cell that has 4 formula units of $Ga_2O_3$ molecule. The main reason of using supercell to model disorder is to explicitly impose translational symmetry over larger length scales so that Bloch waves can be used for eigenfunction expansion and secondly, large cells reduce the error introduced by self image interaction between the impurities/defects. One caveat of such a technique is the increase in computational complexity associated with the increased system size. In this work, a 160 atom supercell as shown in Fig.~\ref{fig:structure}a is used as a model for the EBS calculation. Based on the previous work \cite{algaovol1} it is known that octahedrally coordinated $Ga$ sites are energetically favourable for aluminium $(Al)$ incorporation, so upto $50\%$ $Al$ fraction with respect to all the gallium $(Ga)$ atoms, only the octahedral sites are selected for alloy substitution. Under the above constraints, calculations were performed for different fractions of $Al$ varying from $12.5\%$ to $62.5\%$ with respect to $Ga$ sites. One of the reasons for restricting our calculations upto $62.5\%$ fraction is that the alloy undergoes phase transition beyond this concentration and does not maintain the monoclinic structure of $\beta-Ga_2O_3$ \cite{algao2,algaovol1}.

\section{Special Quasirandom Structures}
The next step is designing the 160 atom supercell in a $2\times2\times2$ configuration of the $AlGaO$ alloy disordered system. One of the standard methods is to enumerate the entire configurational space for a given alloy composition and subsequently averaging the properties for each configuration. Though this method is straightforward, it is computationally very expensive. In this work, \textit{SQS} (Special Quasirandom structures)\cite{sqs1,sqs2} that is generated using the ATAT code\cite{atat2,atat3} is used to overcome this computational bottleneck. The entire premise is to match the site occupancy correlation of different candidate structures to that of the random alloy through permutation where the cost function formulated in terms of correlation mismatch between the candidate structure and the truly random alloy is minimized using Monte-Carlo optimization techniques. In this work, 2 atom and 3 atom clusters with an atomic distance of $5.2$\AA \hspace{0.01cm} and $6.0$\AA \hspace{0.01cm} respectively were used for cluster expansion. The significant reduction of the computational cost that is achieved using SQS shadows the drawback of the small system size where it is very challenging to match the site correlation of the candidate structure to that of the truly random alloy system.

The supercell $(SC)$ and the primitive cell $(PC)$ used for calculation in this study are commensurate \textit{i.e.} the lattice vectors of the supercell and that of the primitive cell are related as $A_i^{SC}=\sum_j\lambda_{ij}a_j^{PC}$ where $\lambda$ is a matrix of integers. Since the $SC$ and $PC$ are commensurate in real space, their Brillouin zones $(BZ)$ are also commensurate in the reciprocal space, as such the reciprocal space points in the supercell BZ $(SCBZ)$ and the primitive cell BZ $(PCBZ)$ satisfy eqn.\ref{eq:1}\cite{bandunfold1,bandunfold2}:

\begin{equation}
    \vec{\textbf{K}} + \vec{\textbf{G}}_{i} = \vec{\textbf{k}}_{i} \quad  i \in [1,N]
    \label{eq:1}
\end{equation}
where N$(N=16)$ is an integer given as $N=\Omega_{SCBZ}/\Omega_{PCBZ}$ \textit{i.e.} the ratio of the volumes of the $BZs$, $\vec{\textbf{K}}$ is the point in the $SCBZ$, $\vec{\textbf{G}}$ is the reciprocal space lattice vector of the $(SC)$ and $\vec{\textbf{k}}$ is the point in the irreducible wedge of the $PCBZ$. The vector representation of eqn.\ref{eq:1} with the primitive and supercell reciprocal space grids represented with blue and red markers respectively is given in Fig.~\ref{fig:structure}b.

\section{Effective Bandstructure (EBS)}

Now for the EBS calculation, as a first step, a map between the reciprocal space points of the $SCBZ$ and $PCBZ$ is created governed by eqn.\ref{eq:1}. Then the self consistent charge density of the supercell disordered system is obtained using VASP\cite{vasp1,vasp2,vasp3} and PAW\cite{paw1} pseudopotential with a planewave cutoff of 600eV on $2\times6\times4$ reciprocal space k-point grid. Then a non-self consistent calculation is done to calculate the eigenvalues and eigenfunctions on the \textit{SCBZ} $\vec{\textbf{K}}$ grid generated using eqn.\ref{eq:1}.  These eigenfunctions are then projected onto the $PCBZ$ grid $(\vec{\textbf{k}})$ to obtain the projection probability which is also known as the spectral weight. The idea of calculating the projection probability lies on the fact that the $PC$ Bloch waves form a complete set and hence the $SC$ eigenfunctions can be expanded in terms of these $PC$ eigenfunctions. This is the crux of the process of unfolding. The spectral weight is given in eqn.\ref{eq:2}:

\begin{equation}
    P_{\vec{\textbf{K}}m}(\vec{\textbf{k}}_i) = \sum_{n}|<\psi_{\vec{\textbf{K}}m}|\psi_{\vec{\textbf{k}}_in}>|^2
    \label{eq:2}
\end{equation}
where, $\vec{\textbf{K}}$ and $\vec{\textbf{k}}$ are in full $SCBZ$ and irreducible wedge of the $PCBZ$ respectively, $m$ and $n$ are the band indices, $\psi$ is the ground state wavefunction and \textbf{P} is the spectral weight. The technique to calculate the spectral weight without the explicit knowledge of the $PC$ wavefunctions is theorized by Zunger \textit{et.al}\cite{bandunfold1,bandunfold2} and the expression is given in the supplementary material. Spectral weight (\textbf{P}) gives a quantitative insight into the Bloch character that is preserved in the supercell when there is no translational symmetry in the crystal; due to disorder introduced in the bulk $\beta-Ga_2O_3$ crystal as a result of aluminium incorporation.

This spectral weight is then used to calculate the spectral function as given in eqn.\ref{eq:3}:
\begin{equation}
    A(\vec{\textbf{k}}_i;\epsilon) = \sum_{m}P_{\vec{\textbf{K}}m}\delta({\epsilon - E_{\vec{\textbf{K}}m}})
    \label{eq:3}
\end{equation}
where, $E_{\vec{\textbf{K}}m}$ is the eigenvalue of the supercell in the state given by wavevector $\vec{\textbf{K}}$ and band index $m$ , $\epsilon$ is a uniform energy grid with a grid spacing of $0.1eV$. The delta function $(\delta)$ in eqn.\ref{eq:3} is approximated using a Gaussian with $25meV$ broadening. Using this spectral function, the cumulative sum is evaluated to obtain the EBS as described by Zunger in \cite{bandunfold1,bandunfold2}.

\begin{figure*}[h]
    \includegraphics[width=0.85\linewidth]{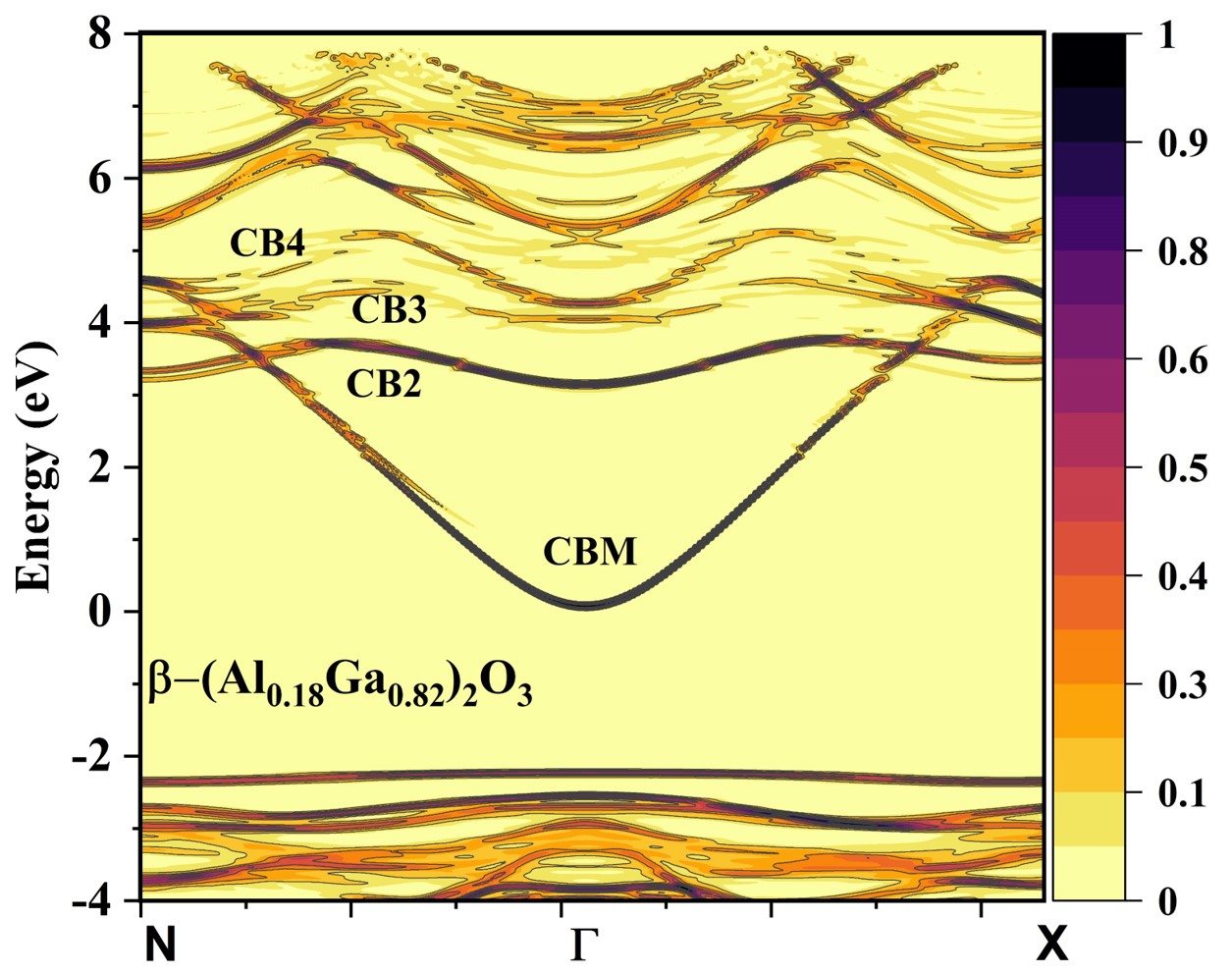}
    \caption{\label{fig:NGX} The \textit{EBS} of $AlGaO$ alloy system with $18.75\%$ aluminium \textit{i.e.} $(Al_{0.18}Ga_{0.82})_2O_3$ where the conduction band minima is set to 0eV and the $BZ$ direction of $N$,$\Gamma$ and $X$ correspond to [0.0 0.5 0.0],[0.0 0.0 0.0] and [0.266 0.266 0]\cite{bz} respectively}
\end{figure*}

\begin{figure*}
    \includegraphics[width=0.9\linewidth]{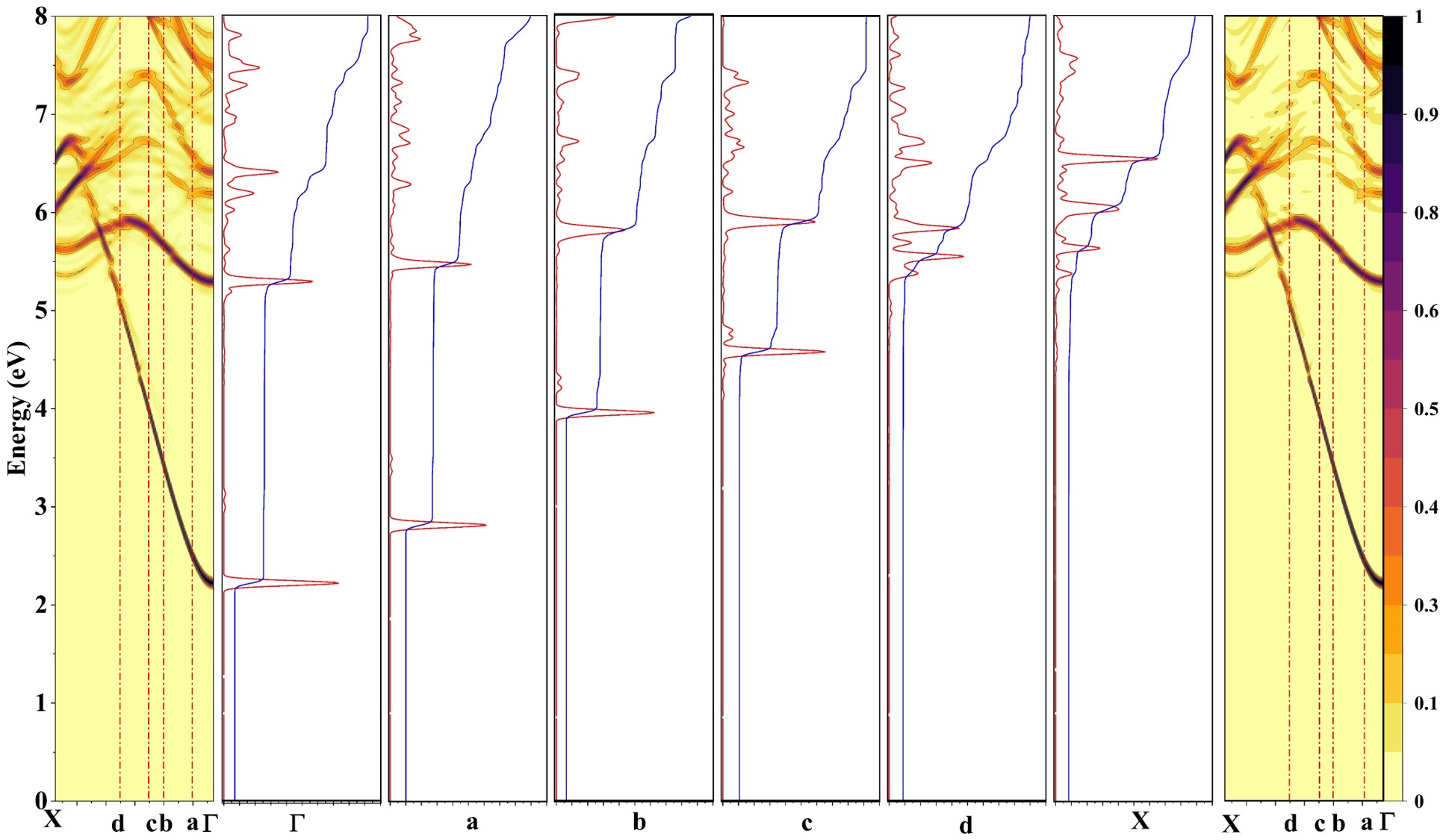}
    \caption{\label{fig:sf_pj_gx}The combined plot of the spectral function represented in \textit{red}, the cumulative sum as \textit{blue}. The presence of states is shown by the peak in the $\delta$ function and the jump in the cumulative sum .The plot on the right and left is the $EBS$ for $(Al_{0.18}Ga_{0.82})_2O_3$ in the $\Gamma$-$X$ direction of the $PCBZ$ with the top of the valence band shifted to $0eV$. Points $a, b, c, d$ represent distinct points in the $\Gamma$-$X$ direction }
\end{figure*}

\begin{figure*}
    \includegraphics[width=0.9\linewidth]{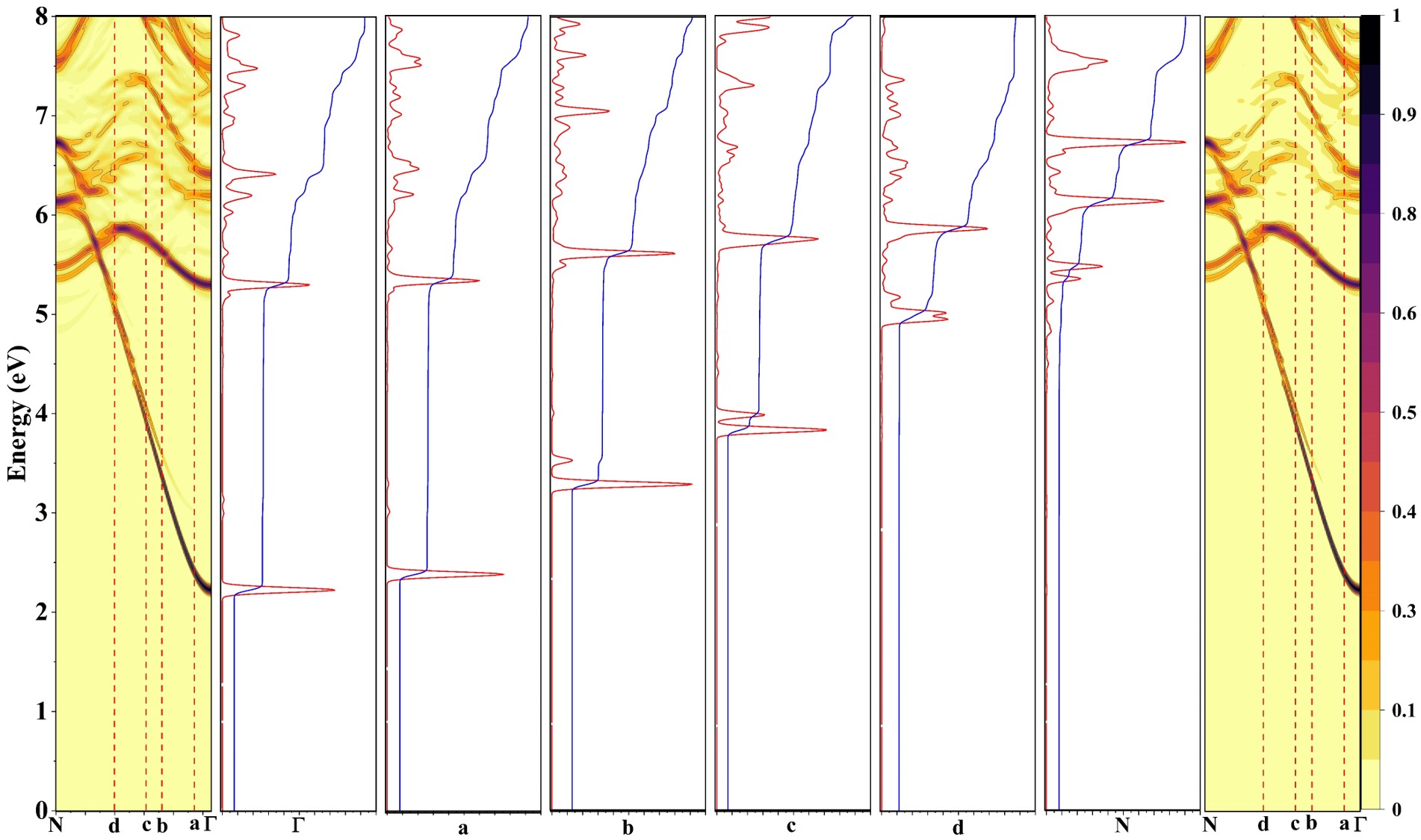}
    \caption{\label{fig:sf_pj_ng}The combined plot of the spectral function represented in \textit{red}, the cumulative sum as \textit{blue}. The presence of states is shown by the peak in the $\delta$ function and the jump in the cumulative sum .The plot on the right and left is the $EBS$ for $(Al_{0.18}Ga_{0.82})_2O_3$ in the $N$-$\Gamma$ direction of the $PCBZ$ with the top of the valence band shifted to $0eV$. Points $a, b, c, d$ represent distinct points in the $N$-$\Gamma$ direction }
\end{figure*}

The plot of the EBS calculated using the cumulative sum for $Al$ fraction of $x=0.18$ is shown in Fig.~\ref{fig:NGX} along the $N-\Gamma-X$\cite{bz} direction of the irreducible wedge of the $PCBZ$. Here the conduction band minima is set as a reference at $0 eV$. The intensity at each of the $PC$ $\vec{\textbf{k}}$ point along irreducible BZ directions gives a qualitative measurement of the amount of Bloch character that is preserved at the equivalent $SC$ $\vec{\textbf{K}}$ related by eqn.\ref{eq:1} and also the presence of an valid energy state. In order to further develop insight of the effective band structure, a plot of the spectral function in the $\Gamma-X$ and $N-\Gamma$ direction is presented in Fig.~\ref{fig:sf_pj_gx} and Fig.~\ref{fig:sf_pj_ng} respectively, which provides 2 different ways to visualize the spectral function. The plot in red labeled $\Gamma,a,...,X$ and $\Gamma,a,...,N$ in the two figures is the spectral function at each $\vec{\textbf{k}}$ point given by eqn.\ref{eq:2} whereas the plot in blue in the same figures shows the cumulative sum of the spectral function. The spectral function are at discrete $\vec{\textbf{k}}$ points in the $PCBZ$ marked by vertical red lines in the $EBS$ in the $\Gamma-X$ and $N-\Gamma$ direction in left and rightmost plots in Fig.~\ref{fig:sf_pj_gx} and Fig.~\ref{fig:sf_pj_ng} . At each of these $\vec{\textbf{k}}$ points, the existence of peaks (red) represent the presence of a state in the $PC$ which is also depicted as a jump in the cumulative sum (blue curve), whereas the relative magnitude of the peak gives a qualitative estimate of the preserved Bloch character at reciprocal point $\vec{\textbf{K}}$ in the supercell.

The isolated sharp peaks near the conduction band minima (CBM) in Fig.~\ref{fig:sf_pj_gx} and Fig.~\ref{fig:sf_pj_ng} represent that the band is fairly unaffected by the disorder introduced due to aluminium incorporation and retains its \textit{s}-orbital type character mainly contributed by the $Ga$ atoms. Moving higher up in energies along the Brillouin zone direction, multiple peaks can be seen clustered in a given energy window as shown in Fig.~\ref{fig:sf_pj_gx}b,c,d,X and Fig.~\ref{fig:sf_pj_ng}b,c,d,N starting around $3.5eV$ energy. This clustering of peaks is the direct result of the disorder and is known as band broadening. The broadening of the bands can be related to the electron lifetime through the moment of the spectral function at each $\vec{\textbf{k}}$ point. Such an approach has been previously used in $AlGaN$ to evaluate alloy scattering rate and the electron mobility\cite{alloyscat1}. This provides an alternative to the currently used method of using conduction band offset as the scattering potential for the evaluation of the alloy scattering rate.

Another observation that can be made regarding the band broadening is that it is anisotropic as can be seen in Fig.~\ref{fig:NGX} at higher energies in the $1^{st}$ conduction band in the $\Gamma-N$ and $\Gamma-X$ directions of the $PCBZ$. Hence, under high electric fields, the anisotropy in the electron mobility will potentially be further impacted by this disorder induced band broadening. Bands that are much higher in energy \textit{i.e.} $CB2, CB3, CB4$ conduction band are very diffused with much low intensity as can be seen in Fig.~\ref{fig:NGX} and as low magnitude peaks in Fig.~\ref{fig:sf_pj_gx} and Fig.~\ref{fig:sf_pj_ng} which are closely clustered representing significant band mixing and disorder induced broadening as a result of loss of translational symmetry. On the other hand the bottom of the $1^{st}$ conduction band (low energy states) is fairly isotropic and the anisotropy in the low field electron mobility will mainly be due to the anisotorpic phonon polarization as in bulk $\beta-Ga_2O_3$\cite{aniso1}. The top of the valence band remains mostly flat contributed by oxygen $p$-orbitals, but there is significant broadening in bands deep in the valence band energy range as shown in Fig.~\ref{fig:NGX}.

\begin{figure*}[!t]
    \centering
    \begin{tabular}{|c|c|}
    \hline
    a).\includegraphics[width=0.5\linewidth]{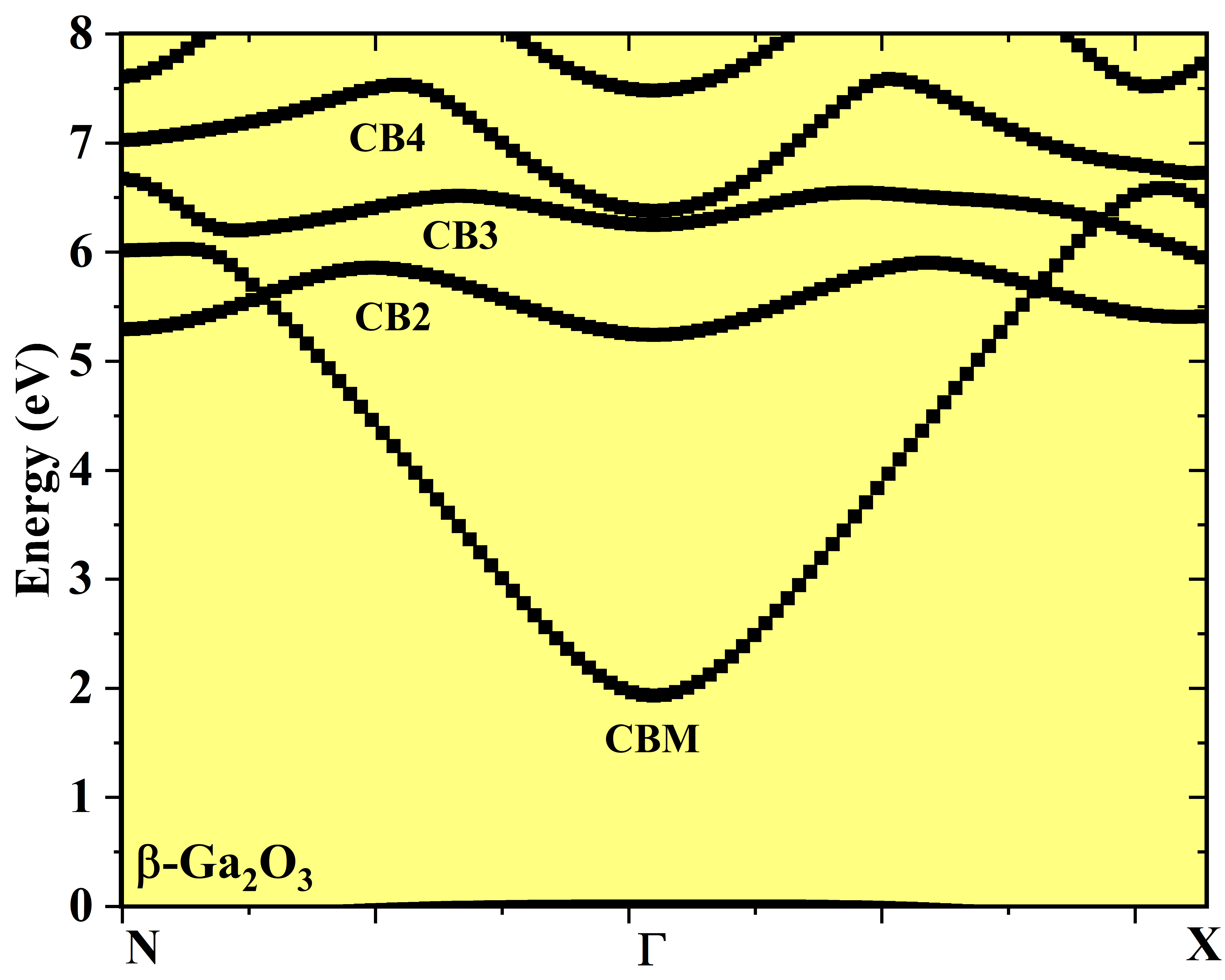} & b).\includegraphics[width=0.5\linewidth]{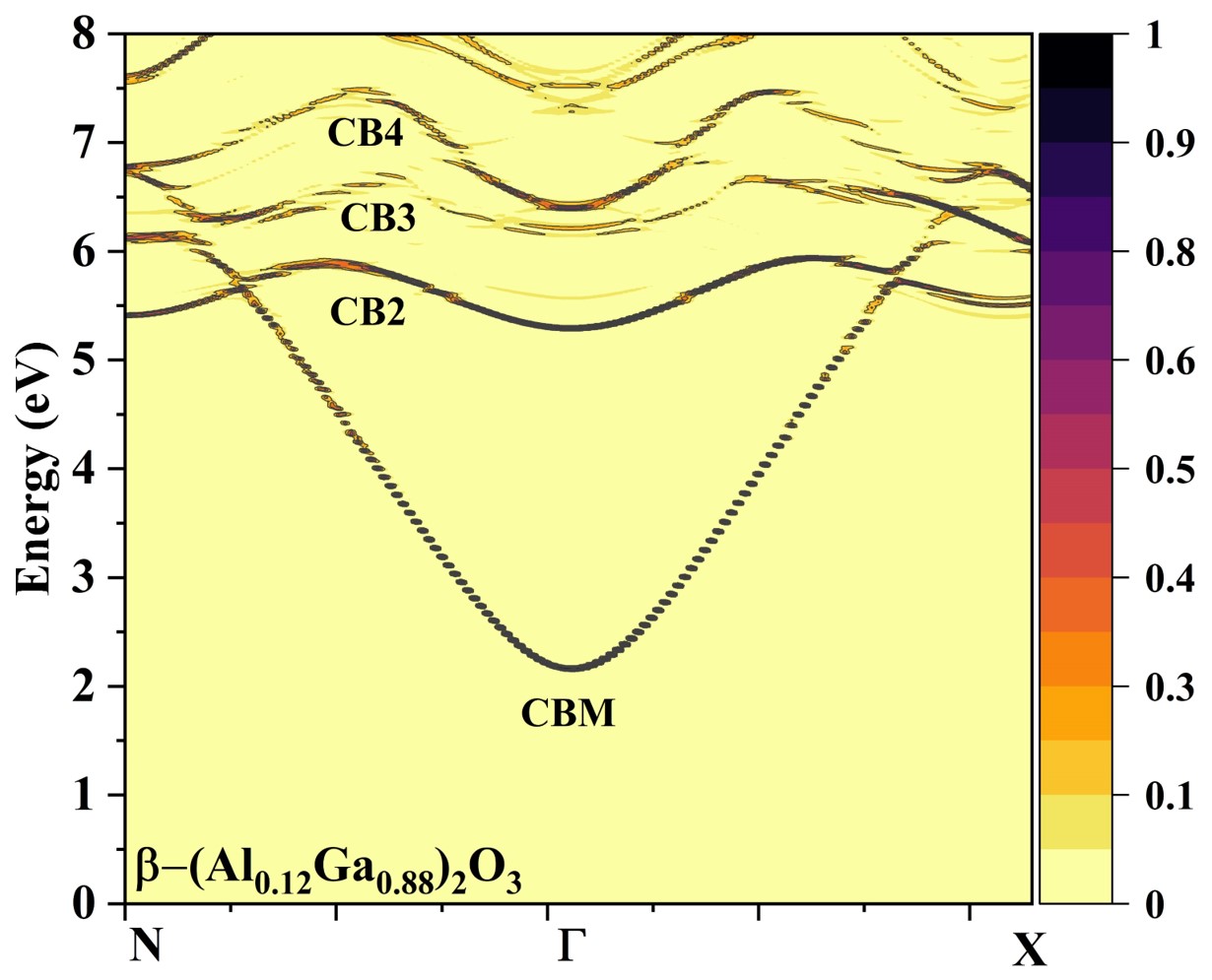}\\
    \hline
    c).\includegraphics[width=0.5\linewidth]{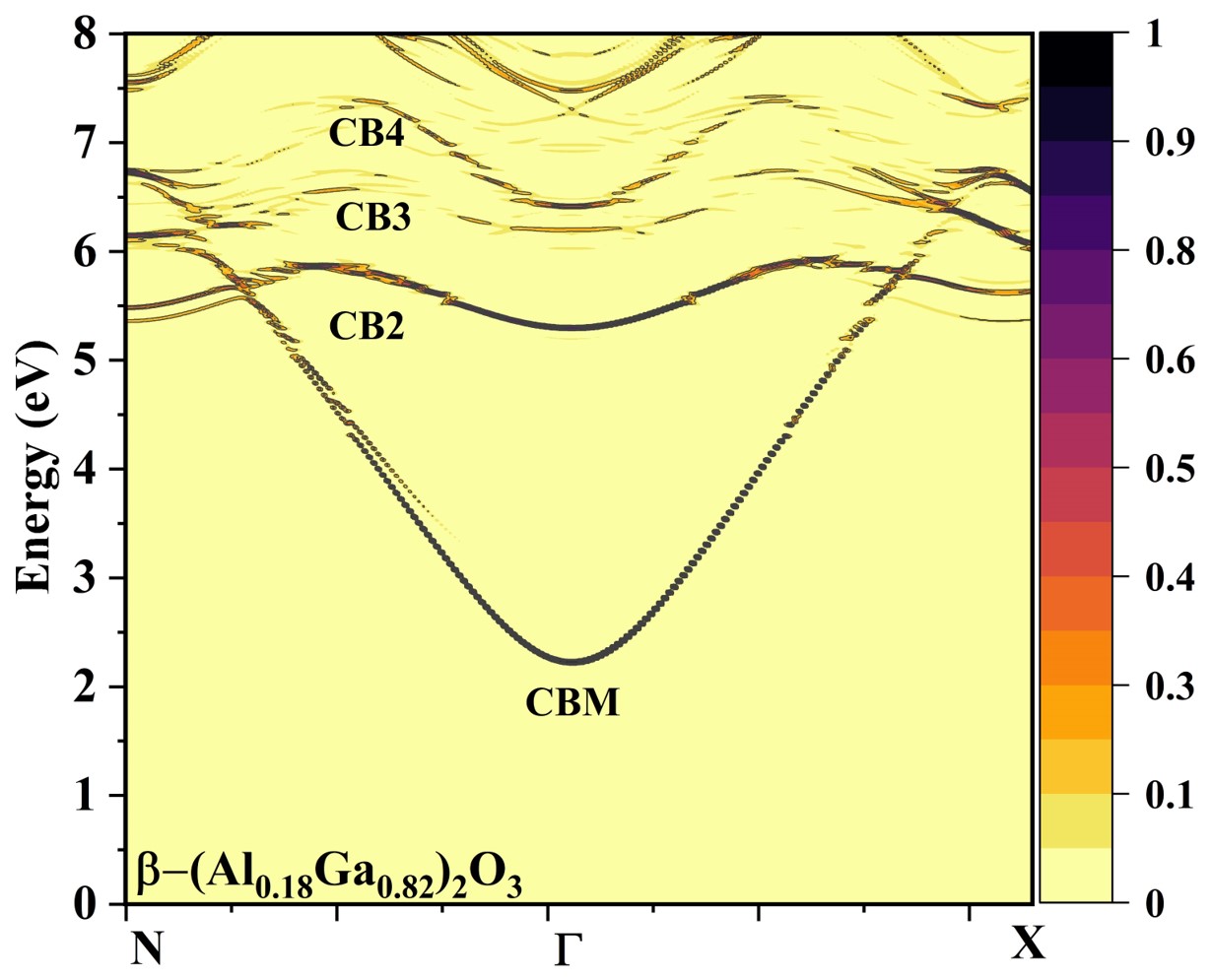} & d).\includegraphics[width=0.5\linewidth]{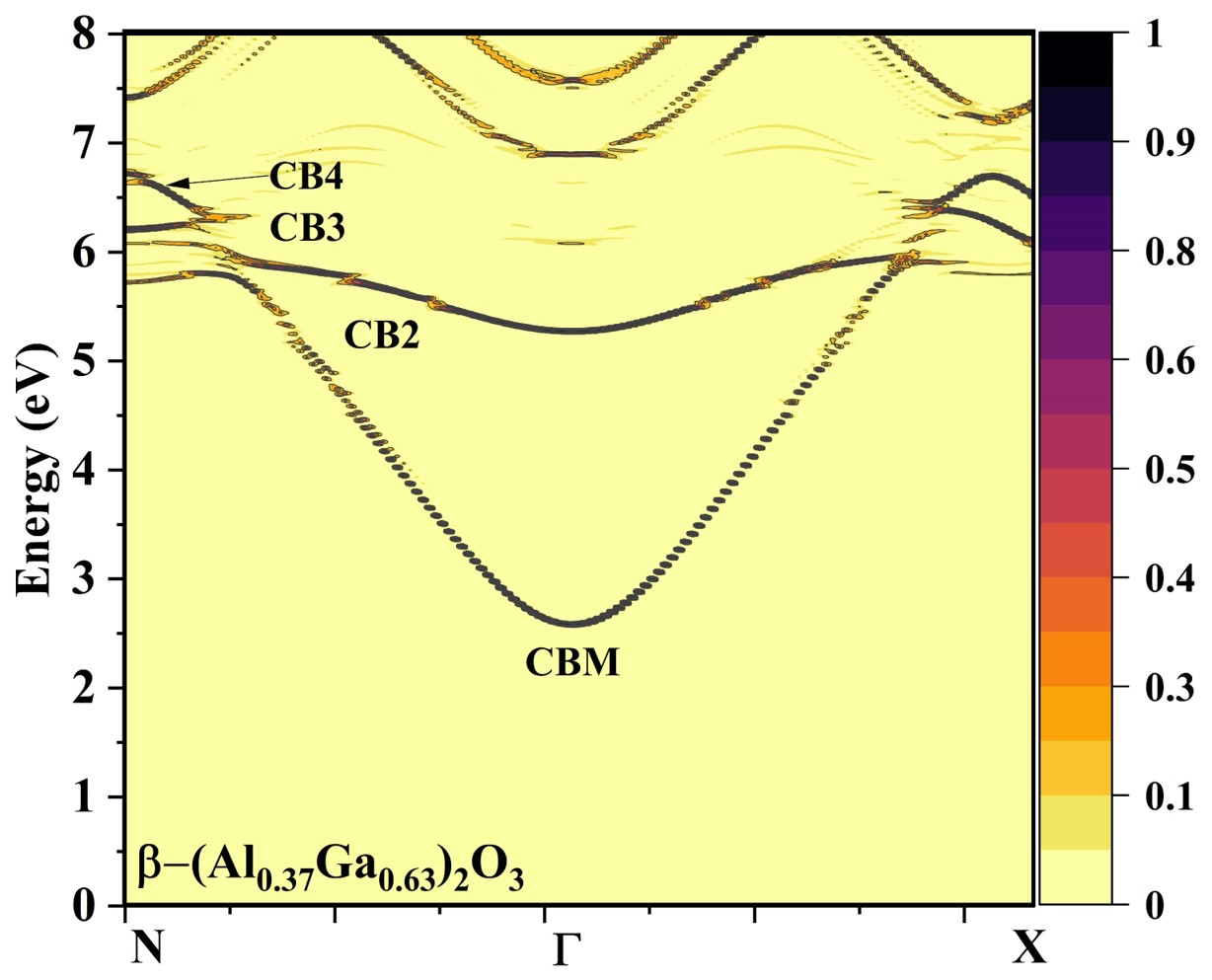}\\
    \hline
    e).\includegraphics[width=0.5\linewidth]{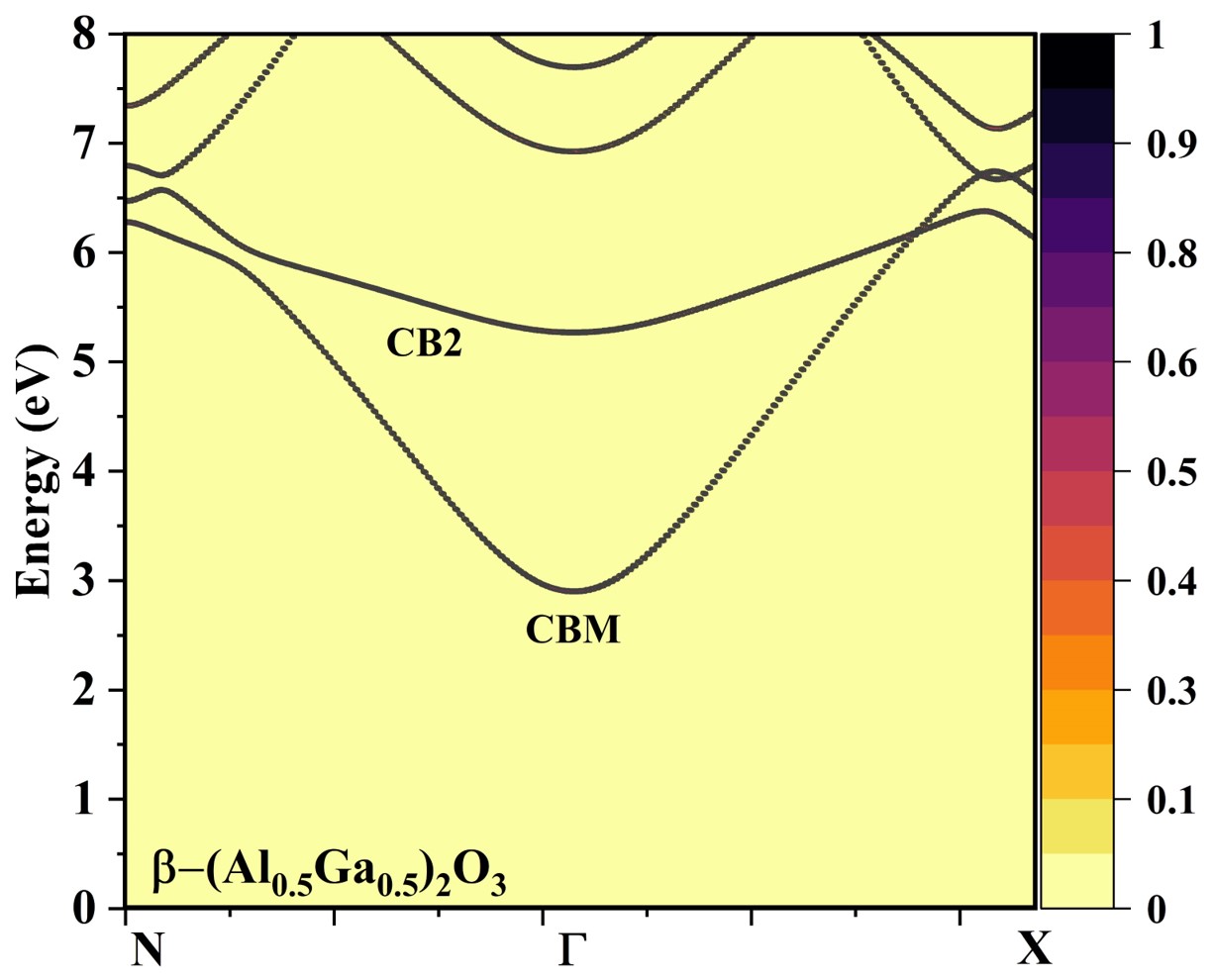} & f).\includegraphics[width=0.5\linewidth]{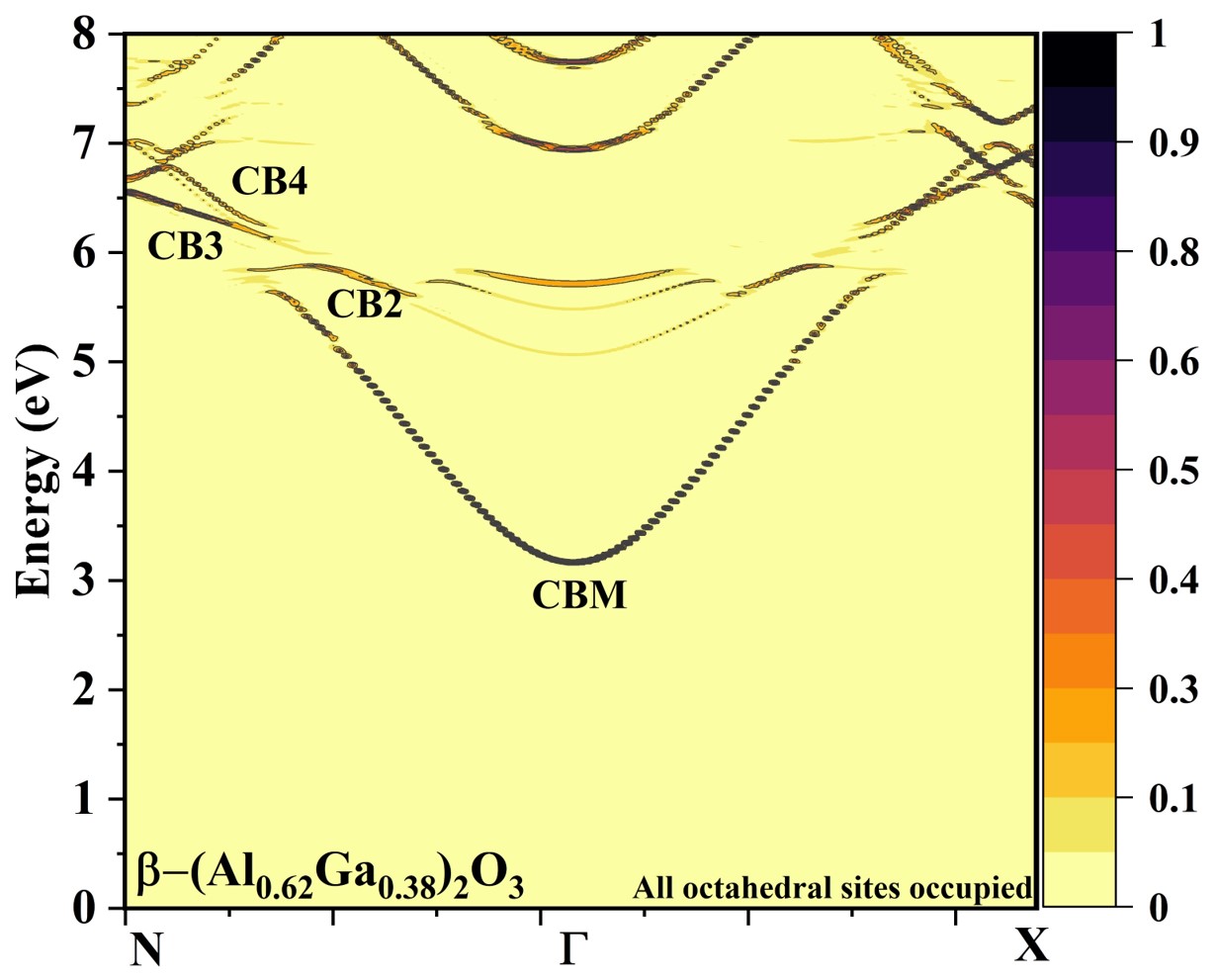}\\
    \hline
    \end{tabular}
    \caption{\label{fig:bandgap}}{The variation of bandgap with \textit{Al} fraction in \textit{AlGaO} alloy; 0\textit{eV} is set to the top of the valence band}
\end{figure*}

\begin{figure*}
    \centering
    \includegraphics[width=0.8\linewidth]{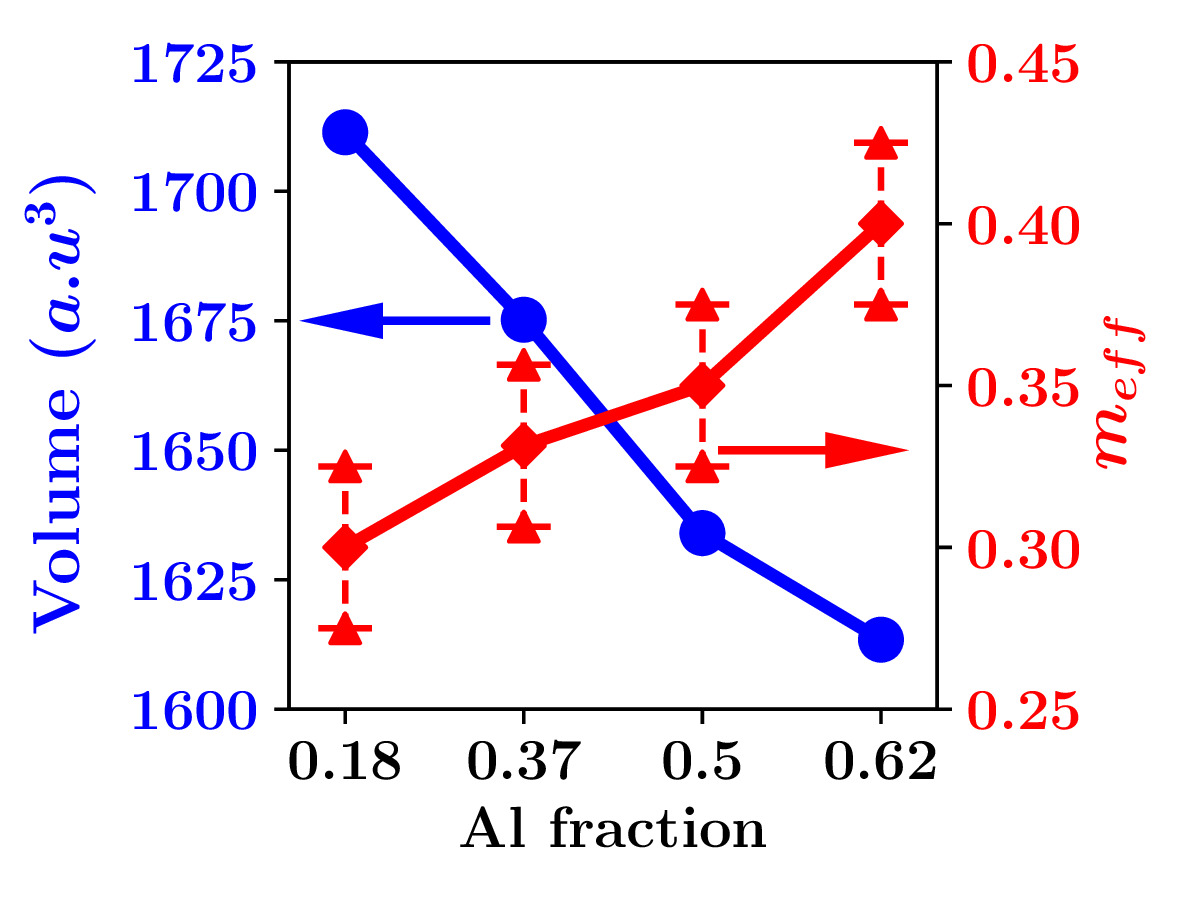}
    \caption{\label{fig:effmass_vol} The variation of volume with \textit{Al} fraction in \textit{blue} and the electron effective mass in \textit{red}. Error bar represents the numerical error from regression}
\end{figure*}

Next, the effect of varying $Al$ fraction on the electron effective mass, volume and bandgap is analyzed as shown in Fig.~\ref{fig:effmass_vol} and figures in Table.~\ref{fig:bandgap} respectively. It can be seen from Fig.~\ref{fig:effmass_vol} that with the increase in $Al$ concentration, the volume of the supercell decreases because the radius of the $Al$ atom is smaller than the $Ga$ atoms, as a result the crystal contracts. The variation of volume follows Vegard's law but at $50\%$ $Al$ fraction, there is change in the slope of the curve, as after all the octahedral sites are exhausted, the tetrahedral sites begin to be substituted by aluminium. This behaviour is also observed in previous theoretical studies\cite{algaovol1}. On the other hand, the effective mass increases with the increase in $Al$ concentration. Based on this observation, a first order conclusion can be reached that the low field electron mobility will potentially decrease with increasing $Al$ fraction in the $AlGaO$ alloy system. The bandgap also follows a trend similar to that of the effective mass, where pristine $\beta-Ga_2O_3$ is at the lower end of the gap and $Al_2O_3$ having the highest gap. Figures in Table.~\ref{fig:bandgap} shows the variation of bandgap of $AlGaO$ alloys and also the change in bands with increasing $Al$ concentration in the alloy. The self consistent DFT computations are performed under the GGA$(PBE\cite{pbe})$ approximation to minimize the increase in the computational complexity due to the large system size. As such, the bandgap is underestimated for all fractions of aluminium in the alloy largely due to the error introduced by the electron self interaction. To ascertain the effect of using GGA approximation rather than the higher order hybrid functional approximation on the electronic dispersion, we run a HSE06 functional calculation on a smaller 80 atom supercell to balance the computational cost and the system size. The EBS for the hybrid functional is shown in the {supplementary material}. Apart from the obvious increase in the bandgap, there is no change in the band topology with the change in the exchange-correlation functional.

The impact of increasing disorder on the electron dispersion can be clearly seen in the figures in Table.~\ref{fig:bandgap}. Concentrating our analysis to the $CB2$ conduction band, for $12.5\%$ aluminium fraction in figure Table.~\ref{fig:bandgap}b , the band appears unaffected except near the $BZ$ edge where there are signatures of broadening as the band become more diffused. Upon futher increasing the aluminium fraction, the disorder induced broadening increases as seen in figure Table.~\ref{fig:bandgap}d where the $CB2$ conduction band is more diffused for $37.5\%$ when compared to $18.75\%$ in figure Table.~\ref{fig:bandgap}c and $12.5\%$ aluminium in figure Table.~\ref{fig:bandgap}b. As for CB3 and CB4, their signature becomes weak as $Al$ fraction increases in the alloy and completely vanish at 50\% $Al$ fraction. At $50\%$ aluminium fraction in figure Table.~\ref{fig:bandgap}e the structure obtains full order with all the octahedral sites occupied by aluminium and the tetrahedral sited by gallium. The structure at this concentration of aluminium has no local disorder which results in a broadening free dispersion. The band again becomes diffused when the aluminium fraction is further increased as shown in figure Table.~\ref{fig:bandgap}f for $62.5\%$ aluminium fraction especially CB2 and above. This observation can also be extended to the $1^{st}$ conduction band at higher energies where the minima remains unaffected due to the disorder.

\begin{figure*}[!t]
    \centering
    \begin{tabular}{|c|c|}
    \hline
    \includegraphics[width=0.5\linewidth]{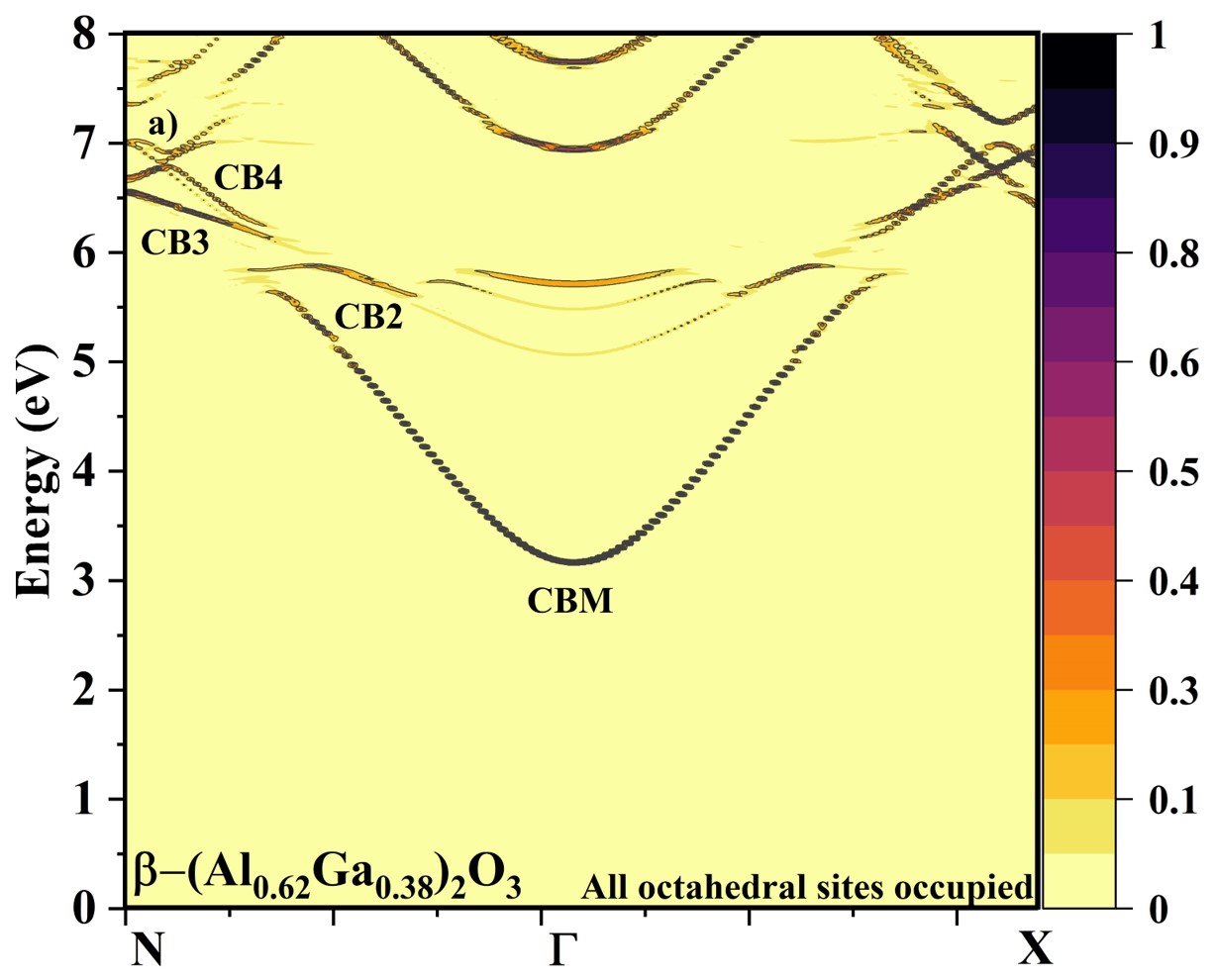} & \includegraphics[width=0.5\linewidth]{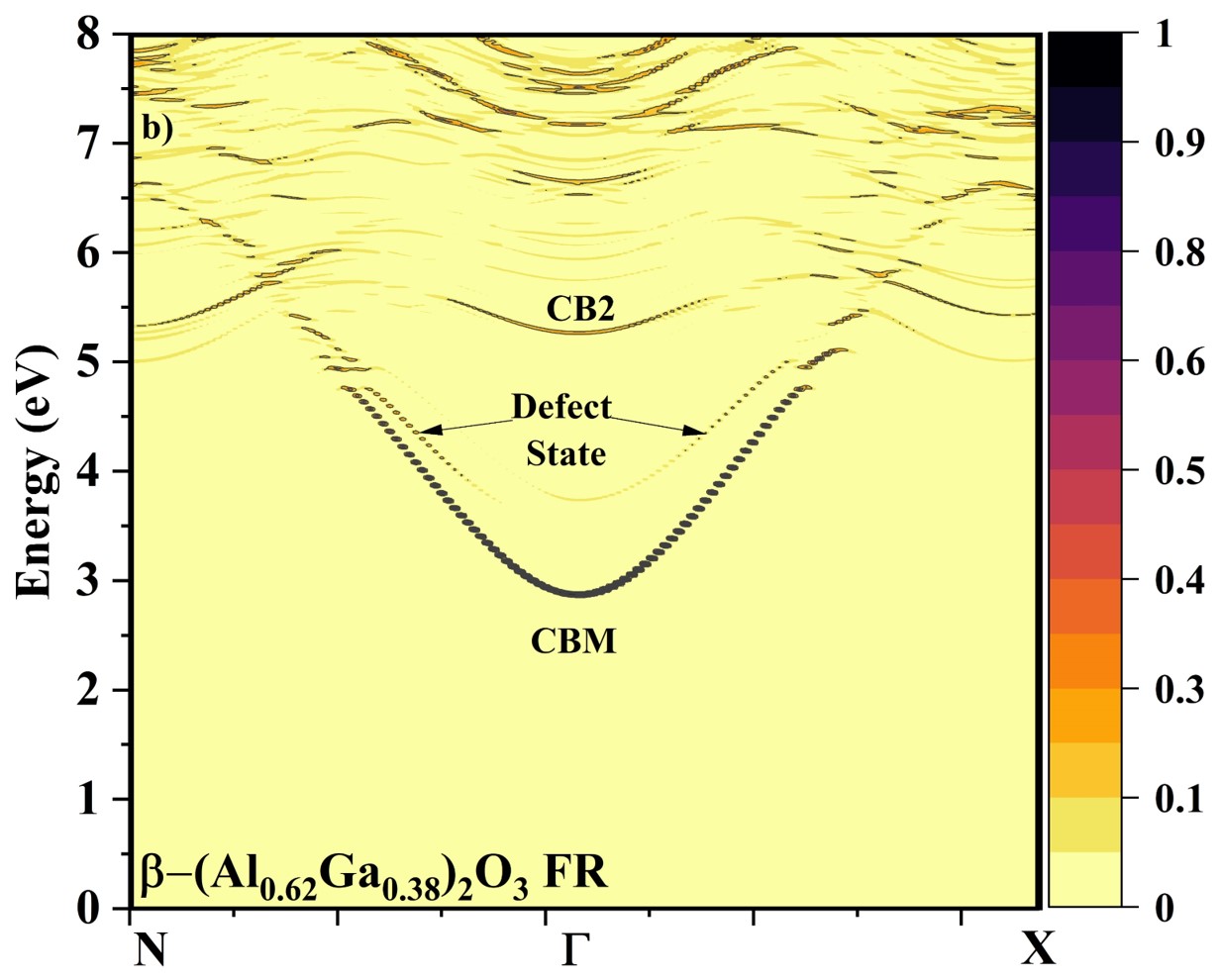}\\
    \hline
    \end{tabular}
    \caption{\label{fig:disorder}}{The EBS of the partially ordered structure in (a) and the fully disordered structure in (b) where FR stands for full random. The EBS for the fully disordered structure shows significant band broadening in the higher conduction band \textit{CB2} and higher which will potentially have an impact on high field electron mobility}
\end{figure*}

To further gain an understanding into how disorder affects the electronic band topology, we calculate the EBS for two AlGaO supercell structures with $62.5\%$ aluminium fraction with partial and full disorder respectively. For partial disorder, as stated earlier, the aluminium site occupancy is energetically constrained to the octahedral gallium sites, whereas no such constraint is imposed for the fully disordered structure where both the octahedral and tetrahedral sites have equal probability of aluminium incorporation. The figures in Table.~\ref{fig:disorder}a and b shows the calculated EBS for the two cases. As expected, the bottom of the conduction band still remains isolated and unaffected due to the break in the translational symmetry, but the higher energy bands for the fully disordered case is much more diffused as compared to that of the partially disordered case which leads us to believe that the low field mobility will not be affected much due to the alloy induced disorder where the majority electron population is very close to the conduction band minima, but the same cannot be said for high field mobility where the band broadening and its diffused nature is directly proportional to the electron lifetime and hence the mobility. It can be seen from the figure in Table.~\ref{fig:disorder}b the appearance of the defect states above the $1^{st}$ conduction band which is attributed to the disorder in the structure. The higher conduction bands CB2 and above for a fully disordered case is much more diffused with more number of spectral function peaks clustered within a given energy window making the identification of peaks and thus the presentation of EBS a challenge.

In conclusion, the electronic band structure of disordered $AlGaO$ alloy system was calculated using the technique of \textit{band-unfolding} employing commensurate supercells to model the impurities/defects. A qualitative trend in the bandgap and electron effective mass with respect to aluminium fraction was also obtained. A common observation among different configurations is that the lowermost conduction band near the band minima remains fairly unaffected by the disorder hence Vegard's law could be used for low-field transport calculations. Whereas there is significant impact on higher conduction bands (broadening) potentially affecting the high field electron transport properties. The technique of \textit{band-unfolding} further helps in extracting information that is obscured in the $(SCBZ)$ electron dispersion due to significantly more orbitals clustered within a given energy window. This technique provides an insight into the variation of electron effective mass and can be used to model electron transport mechanism in disordered structures.

See the supplementary material for $k$-point grid convergence and the effect of strain on the $EBS$. 
\begin{acknowledgments}
We acknowledge the support from AFOSR (Air Force Office of Scientific Research) under award FA9550-18-1-0479 (Program Manager: Ali Sayir), from NSF under award  ECCS 2019749, and the Center for Computational Research (CCR) at University at Buffalo.
\end{acknowledgments}

\section*{DATA AVAILABILITY}
The data that support the findings of this study are available from the corresponding author upon request.

\section*{REFERENCES}
\nocite{*}
\bibliography{references}

\providecommand{\noopsort}[1]{}\providecommand{\singleletter}[1]{#1}%
\begin{thebibliography}{45}%
\makeatletter
\providecommand \@ifxundefined [1]{%
 \@ifx{#1\undefined}
}%
\providecommand \@ifnum [1]{%
 \ifnum #1\expandafter \@firstoftwo
 \else \expandafter \@secondoftwo
 \fi
}%
\providecommand \@ifx [1]{%
 \ifx #1\expandafter \@firstoftwo
 \else \expandafter \@secondoftwo
 \fi
}%
\providecommand \natexlab [1]{#1}%
\providecommand \enquote  [1]{``#1''}%
\providecommand \bibnamefont  [1]{#1}%
\providecommand \bibfnamefont [1]{#1}%
\providecommand \citenamefont [1]{#1}%
\providecommand \href@noop [0]{\@secondoftwo}%
\providecommand \href [0]{\begingroup \@sanitize@url \@href}%
\providecommand \@href[1]{\@@startlink{#1}\@@href}%
\providecommand \@@href[1]{\endgroup#1\@@endlink}%
\providecommand \@sanitize@url [0]{\catcode `\\12\catcode `\$12\catcode
  `\&12\catcode `\#12\catcode `\^12\catcode `\_12\catcode `\%12\relax}%
\providecommand \@@startlink[1]{}%
\providecommand \@@endlink[0]{}%
\providecommand \url  [0]{\begingroup\@sanitize@url \@url }%
\providecommand \@url [1]{\endgroup\@href {#1}{\urlprefix }}%
\providecommand \urlprefix  [0]{URL }%
\providecommand \Eprint [0]{\href }%
\providecommand \doibase [0]{http://dx.doi.org/}%
\providecommand \selectlanguage [0]{\@gobble}%
\providecommand \bibinfo  [0]{\@secondoftwo}%
\providecommand \bibfield  [0]{\@secondoftwo}%
\providecommand \translation [1]{[#1]}%
\providecommand \BibitemOpen [0]{}%
\providecommand \bibitemStop [0]{}%
\providecommand \bibitemNoStop [0]{.\EOS\space}%
\providecommand \EOS [0]{\spacefactor3000\relax}%
\providecommand \BibitemShut  [1]{\csname bibitem#1\endcsname}%
\let\auto@bib@innerbib\@empty
\bibitem [{\citenamefont {Roy}, \citenamefont {Hill},\ and\ \citenamefont
  {Osborn}(1952)}]{poly1}%
  \BibitemOpen
  \bibfield  {author} {\bibinfo {author} {\bibfnamefont {R.}~\bibnamefont
  {Roy}}, \bibinfo {author} {\bibfnamefont {V.~G.}\ \bibnamefont {Hill}}, \
  and\ \bibinfo {author} {\bibfnamefont {E.~F.}\ \bibnamefont {Osborn}},\
  }\bibfield  {title} {\enquote {\bibinfo {title} {Polymorphism of
  ${Ga}_{2}{O}_{3}$ and the system ${Ga}_{2}{O}_{3}-{H}_{2}{O}$},}\ }\href
  {\doibase 10.1021/ja01123a039} {\bibfield  {journal} {\bibinfo  {journal}
  {Journal of the American Chemical Society}\ }\textbf {\bibinfo {volume}
  {74}},\ \bibinfo {pages} {719--722} (\bibinfo {year} {1952})}\BibitemShut
  {NoStop}%
\bibitem [{\citenamefont {Roy}, \citenamefont {Hill},\ and\ \citenamefont
  {Osborn}(1953)}]{thermstb1}%
  \BibitemOpen
  \bibfield  {author} {\bibinfo {author} {\bibfnamefont {R.}~\bibnamefont
  {Roy}}, \bibinfo {author} {\bibfnamefont {V.~G.}\ \bibnamefont {Hill}}, \
  and\ \bibinfo {author} {\bibfnamefont {E.~F.}\ \bibnamefont {Osborn}},\
  }\bibfield  {title} {\enquote {\bibinfo {title} {Polymorphs of alumina and
  gallia},}\ }\href {\doibase 10.1021/ie50520a047} {\bibfield  {journal}
  {\bibinfo  {journal} {Industrial \& Engineering Chemistry}\ }\textbf
  {\bibinfo {volume} {45}},\ \bibinfo {pages} {819--820} (\bibinfo {year}
  {1953})},\ \Eprint {http://arxiv.org/abs/https://doi.org/10.1021/ie50520a047}
  {https://doi.org/10.1021/ie50520a047} \BibitemShut {NoStop}%
\bibitem [{\citenamefont {Razeghi}\ \emph {et~al.}(2018)\citenamefont
  {Razeghi}, \citenamefont {Park}, \citenamefont {McClintock}, \citenamefont
  {Pavlidis}, \citenamefont {Teherani}, \citenamefont {Rogers}, \citenamefont
  {Magill}, \citenamefont {Khodaparast}, \citenamefont {Xu}, \citenamefont
  {Wu},\ and\ \citenamefont {Dravid}}]{thermstb2}%
  \BibitemOpen
  \bibfield  {author} {\bibinfo {author} {\bibfnamefont {M.}~\bibnamefont
  {Razeghi}}, \bibinfo {author} {\bibfnamefont {J.-H.}\ \bibnamefont {Park}},
  \bibinfo {author} {\bibfnamefont {R.}~\bibnamefont {McClintock}}, \bibinfo
  {author} {\bibfnamefont {D.}~\bibnamefont {Pavlidis}}, \bibinfo {author}
  {\bibfnamefont {F.~H.}\ \bibnamefont {Teherani}}, \bibinfo {author}
  {\bibfnamefont {D.~J.}\ \bibnamefont {Rogers}}, \bibinfo {author}
  {\bibfnamefont {B.~A.}\ \bibnamefont {Magill}}, \bibinfo {author}
  {\bibfnamefont {G.~A.}\ \bibnamefont {Khodaparast}}, \bibinfo {author}
  {\bibfnamefont {Y.}~\bibnamefont {Xu}}, \bibinfo {author} {\bibfnamefont
  {J.}~\bibnamefont {Wu}}, \ and\ \bibinfo {author} {\bibfnamefont {V.~P.}\
  \bibnamefont {Dravid}},\ }\bibfield  {title} {\enquote {\bibinfo {title} {{A
  review of the growth, doping, and applications of $\beta-{Ga}_{2}{O}_{3}$
  thin films}},}\ }in\ \href {\doibase 10.1117/12.2302471} {\emph {\bibinfo
  {booktitle} {Oxide-based Materials and Devices IX}}},\ Vol.\ \bibinfo
  {volume} {10533},\ \bibinfo {editor} {edited by\ \bibinfo {editor}
  {\bibfnamefont {D.~J.}\ \bibnamefont {Rogers}}, \bibinfo {editor}
  {\bibfnamefont {D.~C.}\ \bibnamefont {Look}}, \ and\ \bibinfo {editor}
  {\bibfnamefont {F.~H.}\ \bibnamefont {Teherani}}},\ \bibinfo {organization}
  {International Society for Optics and Photonics}\ (\bibinfo  {publisher}
  {SPIE},\ \bibinfo {year} {2018})\ pp.\ \bibinfo {pages} {21 --
  44}\BibitemShut {NoStop}%
\bibitem [{\citenamefont {Tippins}(1965)}]{gap1}%
  \BibitemOpen
  \bibfield  {author} {\bibinfo {author} {\bibfnamefont {H.~H.}\ \bibnamefont
  {Tippins}},\ }\bibfield  {title} {\enquote {\bibinfo {title} {Optical
  {A}bsorption and {P}hotoconductivity in the {B}and {E}dge of
  $\beta-{Ga}_{2}{O}{3}$},}\ }\href {\doibase 10.1103/PhysRev.140.A316}
  {\bibfield  {journal} {\bibinfo  {journal} {Phys. Rev.}\ }\textbf {\bibinfo
  {volume} {140}},\ \bibinfo {pages} {A316--A319} (\bibinfo {year}
  {1965})}\BibitemShut {NoStop}%
\bibitem [{\citenamefont {Orita}\ \emph {et~al.}(2000)\citenamefont {Orita},
  \citenamefont {Ohta}, \citenamefont {Hirano},\ and\ \citenamefont
  {Hosono}}]{gap2}%
  \BibitemOpen
  \bibfield  {author} {\bibinfo {author} {\bibfnamefont {M.}~\bibnamefont
  {Orita}}, \bibinfo {author} {\bibfnamefont {H.}~\bibnamefont {Ohta}},
  \bibinfo {author} {\bibfnamefont {M.}~\bibnamefont {Hirano}}, \ and\ \bibinfo
  {author} {\bibfnamefont {H.}~\bibnamefont {Hosono}},\ }\bibfield  {title}
  {\enquote {\bibinfo {title} {Deep-ultraviolet transparent conductive
  $\beta-{Ga}_{2}{O}_{3}$ thin films},}\ }\href {\doibase 10.1063/1.1330559}
  {\bibfield  {journal} {\bibinfo  {journal} {Applied Physics Letters}\
  }\textbf {\bibinfo {volume} {77}},\ \bibinfo {pages} {4166--4168} (\bibinfo
  {year} {2000})}\BibitemShut {NoStop}%
\bibitem [{\citenamefont {He}\ \emph {et~al.}(2006)\citenamefont {He},
  \citenamefont {Orlando}, \citenamefont {Blanco}, \citenamefont {Pandey},
  \citenamefont {Amzallag}, \citenamefont {Baraille},\ and\ \citenamefont
  {R\'erat}}]{thermstb}%
  \BibitemOpen
  \bibfield  {author} {\bibinfo {author} {\bibfnamefont {H.}~\bibnamefont
  {He}}, \bibinfo {author} {\bibfnamefont {R.}~\bibnamefont {Orlando}},
  \bibinfo {author} {\bibfnamefont {M.~A.}\ \bibnamefont {Blanco}}, \bibinfo
  {author} {\bibfnamefont {R.}~\bibnamefont {Pandey}}, \bibinfo {author}
  {\bibfnamefont {E.}~\bibnamefont {Amzallag}}, \bibinfo {author}
  {\bibfnamefont {I.}~\bibnamefont {Baraille}}, \ and\ \bibinfo {author}
  {\bibfnamefont {M.}~\bibnamefont {R\'erat}},\ }\bibfield  {title} {\enquote
  {\bibinfo {title} {First-principles study of the structural, electronic, and
  optical properties of ${Ga}_{2}{O}_{3}$ in its monoclinic and hexagonal
  phases},}\ }\href {\doibase 10.1103/PhysRevB.74.195123} {\bibfield  {journal}
  {\bibinfo  {journal} {Phys. Rev. B}\ }\textbf {\bibinfo {volume} {74}},\
  \bibinfo {pages} {195123} (\bibinfo {year} {2006})}\BibitemShut {NoStop}%
\bibitem [{\citenamefont {Higashiwaki}\ \emph {et~al.}(2013)\citenamefont
  {Higashiwaki}, \citenamefont {Sasaki}, \citenamefont {Kamimura},
  \citenamefont {Hoi~Wong}, \citenamefont {Krishnamurthy}, \citenamefont
  {Kuramata}, \citenamefont {Masui},\ and\ \citenamefont
  {Yamakoshi}}]{breakdown1}%
  \BibitemOpen
  \bibfield  {author} {\bibinfo {author} {\bibfnamefont {M.}~\bibnamefont
  {Higashiwaki}}, \bibinfo {author} {\bibfnamefont {K.}~\bibnamefont {Sasaki}},
  \bibinfo {author} {\bibfnamefont {T.}~\bibnamefont {Kamimura}}, \bibinfo
  {author} {\bibfnamefont {M.}~\bibnamefont {Hoi~Wong}}, \bibinfo {author}
  {\bibfnamefont {D.}~\bibnamefont {Krishnamurthy}}, \bibinfo {author}
  {\bibfnamefont {A.}~\bibnamefont {Kuramata}}, \bibinfo {author}
  {\bibfnamefont {T.}~\bibnamefont {Masui}}, \ and\ \bibinfo {author}
  {\bibfnamefont {S.}~\bibnamefont {Yamakoshi}},\ }\bibfield  {title} {\enquote
  {\bibinfo {title} {Depletion-mode ga2o3 metal-oxide-semiconductor
  field-effect transistors on $\beta-{Ga}_{2}{O}_{3}$ (010) substrates and
  temperature dependence of their device characteristics},}\ }\href {\doibase
  10.1063/1.4821858} {\bibfield  {journal} {\bibinfo  {journal} {Applied
  Physics Letters}\ }\textbf {\bibinfo {volume} {103}},\ \bibinfo {pages}
  {123511} (\bibinfo {year} {2013})}\BibitemShut {NoStop}%
\bibitem [{\citenamefont {Ghosh}\ and\ \citenamefont
  {Singisetti}(2017)}]{satvelocity1}%
  \BibitemOpen
  \bibfield  {author} {\bibinfo {author} {\bibfnamefont {K.}~\bibnamefont
  {Ghosh}}\ and\ \bibinfo {author} {\bibfnamefont {U.}~\bibnamefont
  {Singisetti}},\ }\bibfield  {title} {\enquote {\bibinfo {title} {Ab initio
  velocity-field curves in monoclinic $\beta-{Ga}_{2}{O}_{3}$},}\ }\href
  {\doibase 10.1063/1.4986174} {\bibfield  {journal} {\bibinfo  {journal}
  {Journal of Applied Physics}\ }\textbf {\bibinfo {volume} {122}},\ \bibinfo
  {pages} {035702} (\bibinfo {year} {2017})}\BibitemShut {NoStop}%
\bibitem [{\citenamefont {Higashiwaki}\ \emph {et~al.}(2014)\citenamefont
  {Higashiwaki}, \citenamefont {Sasaki}, \citenamefont {Kuramata},
  \citenamefont {Masui},\ and\ \citenamefont {Yamakoshi}}]{bfom1}%
  \BibitemOpen
  \bibfield  {author} {\bibinfo {author} {\bibfnamefont {M.}~\bibnamefont
  {Higashiwaki}}, \bibinfo {author} {\bibfnamefont {K.}~\bibnamefont {Sasaki}},
  \bibinfo {author} {\bibfnamefont {A.}~\bibnamefont {Kuramata}}, \bibinfo
  {author} {\bibfnamefont {T.}~\bibnamefont {Masui}}, \ and\ \bibinfo {author}
  {\bibfnamefont {S.}~\bibnamefont {Yamakoshi}},\ }\bibfield  {title} {\enquote
  {\bibinfo {title} {Development of gallium oxide power devices},}\ }\href
  {\doibase https://doi.org/10.1002/pssa.201330197} {\bibfield  {journal}
  {\bibinfo  {journal} {Physica Status Solidi (a)}\ }\textbf {\bibinfo {volume}
  {211}},\ \bibinfo {pages} {21--26} (\bibinfo {year} {2014})}\BibitemShut
  {NoStop}%
\bibitem [{\citenamefont {Higashiwaki}\ \emph {et~al.}(2016)\citenamefont
  {Higashiwaki}, \citenamefont {Sasaki}, \citenamefont {Murakami},
  \citenamefont {Kumagai}, \citenamefont {Koukitu}, \citenamefont {Kuramata},
  \citenamefont {Masui},\ and\ \citenamefont {Yamakoshi}}]{bfom2}%
  \BibitemOpen
  \bibfield  {author} {\bibinfo {author} {\bibfnamefont {M.}~\bibnamefont
  {Higashiwaki}}, \bibinfo {author} {\bibfnamefont {K.}~\bibnamefont {Sasaki}},
  \bibinfo {author} {\bibfnamefont {H.}~\bibnamefont {Murakami}}, \bibinfo
  {author} {\bibfnamefont {Y.}~\bibnamefont {Kumagai}}, \bibinfo {author}
  {\bibfnamefont {A.}~\bibnamefont {Koukitu}}, \bibinfo {author} {\bibfnamefont
  {A.}~\bibnamefont {Kuramata}}, \bibinfo {author} {\bibfnamefont
  {T.}~\bibnamefont {Masui}}, \ and\ \bibinfo {author} {\bibfnamefont
  {S.}~\bibnamefont {Yamakoshi}},\ }\bibfield  {title} {\enquote {\bibinfo
  {title} {Recent progress in ${Ga}_{2}{O}_3$ power devices},}\ }\href
  {\doibase 10.1088/0268-1242/31/3/034001} {\bibfield  {journal} {\bibinfo
  {journal} {Semiconductor Science and Technology}\ }\textbf {\bibinfo {volume}
  {31}},\ \bibinfo {pages} {034001} (\bibinfo {year} {2016})}\BibitemShut
  {NoStop}%
\bibitem [{\citenamefont {Sharma}\ \emph {et~al.}(2020)\citenamefont {Sharma},
  \citenamefont {Zeng}, \citenamefont {Saha},\ and\ \citenamefont
  {Singisetti}}]{brkdwn1}%
  \BibitemOpen
  \bibfield  {author} {\bibinfo {author} {\bibfnamefont {S.}~\bibnamefont
  {Sharma}}, \bibinfo {author} {\bibfnamefont {K.}~\bibnamefont {Zeng}},
  \bibinfo {author} {\bibfnamefont {S.}~\bibnamefont {Saha}}, \ and\ \bibinfo
  {author} {\bibfnamefont {U.}~\bibnamefont {Singisetti}},\ }\bibfield  {title}
  {\enquote {\bibinfo {title} {Field-plated lateral ${Ga}_{2}{O}_{3}$ mosfets
  with polymer passivation and 8.03 kv breakdown voltage},}\ }\href {\doibase
  10.1109/LED.2020.2991146} {\bibfield  {journal} {\bibinfo  {journal} {IEEE
  Electron Device Letters}\ }\textbf {\bibinfo {volume} {41}},\ \bibinfo
  {pages} {836--839} (\bibinfo {year} {2020})}\BibitemShut {NoStop}%
\bibitem [{\citenamefont {Higashiwaki}\ \emph {et~al.}(2012)\citenamefont
  {Higashiwaki}, \citenamefont {Sasaki}, \citenamefont {Kuramata},
  \citenamefont {Masui},\ and\ \citenamefont {Yamakoshi}}]{mesfet1}%
  \BibitemOpen
  \bibfield  {author} {\bibinfo {author} {\bibfnamefont {M.}~\bibnamefont
  {Higashiwaki}}, \bibinfo {author} {\bibfnamefont {K.}~\bibnamefont {Sasaki}},
  \bibinfo {author} {\bibfnamefont {A.}~\bibnamefont {Kuramata}}, \bibinfo
  {author} {\bibfnamefont {T.}~\bibnamefont {Masui}}, \ and\ \bibinfo {author}
  {\bibfnamefont {S.}~\bibnamefont {Yamakoshi}},\ }\bibfield  {title} {\enquote
  {\bibinfo {title} {Gallium oxide ({Ga}2{O}3) metal-semiconductor field-effect
  transistors on single-crystal $\beta-{Ga}_{2}{O}_{3}$ (010) substrates},}\
  }\href {\doibase 10.1063/1.3674287} {\bibfield  {journal} {\bibinfo
  {journal} {Applied Physics Letters}\ }\textbf {\bibinfo {volume} {100}},\
  \bibinfo {pages} {013504} (\bibinfo {year} {2012})}\BibitemShut {NoStop}%
\bibitem [{\citenamefont {Xia}\ \emph {et~al.}(2019)\citenamefont {Xia},
  \citenamefont {Xue}, \citenamefont {Joishi}, \citenamefont {Mcglone},
  \citenamefont {Kalarickal}, \citenamefont {Sohel}, \citenamefont {Brenner},
  \citenamefont {Arehart}, \citenamefont {Ringel}, \citenamefont {Lodha},
  \citenamefont {Lu},\ and\ \citenamefont {Rajan}}]{mesfet2}%
  \BibitemOpen
  \bibfield  {author} {\bibinfo {author} {\bibfnamefont {Z.}~\bibnamefont
  {Xia}}, \bibinfo {author} {\bibfnamefont {H.}~\bibnamefont {Xue}}, \bibinfo
  {author} {\bibfnamefont {C.}~\bibnamefont {Joishi}}, \bibinfo {author}
  {\bibfnamefont {J.}~\bibnamefont {Mcglone}}, \bibinfo {author} {\bibfnamefont
  {N.~K.}\ \bibnamefont {Kalarickal}}, \bibinfo {author} {\bibfnamefont
  {S.~H.}\ \bibnamefont {Sohel}}, \bibinfo {author} {\bibfnamefont
  {M.}~\bibnamefont {Brenner}}, \bibinfo {author} {\bibfnamefont
  {A.}~\bibnamefont {Arehart}}, \bibinfo {author} {\bibfnamefont
  {S.}~\bibnamefont {Ringel}}, \bibinfo {author} {\bibfnamefont
  {S.}~\bibnamefont {Lodha}}, \bibinfo {author} {\bibfnamefont
  {W.}~\bibnamefont {Lu}}, \ and\ \bibinfo {author} {\bibfnamefont
  {S.}~\bibnamefont {Rajan}},\ }\bibfield  {title} {\enquote {\bibinfo {title}
  {$\beta-{Ga}_{2}{O}_{3}$ {D}elta-{D}oped {F}ield-{E}ffect {T}ransistors
  {W}ith {C}urrent {G}ain {C}utoff {F}requency of 27 {G}hz},}\ }\href {\doibase
  10.1109/LED.2019.2920366} {\bibfield  {journal} {\bibinfo  {journal} {IEEE
  Electron Device Letters}\ }\textbf {\bibinfo {volume} {40}},\ \bibinfo
  {pages} {1052--1055} (\bibinfo {year} {2019})}\BibitemShut {NoStop}%
\bibitem [{\citenamefont {Bhattacharyya}\ \emph
  {et~al.}(2021{\natexlab{a}})\citenamefont {Bhattacharyya}, \citenamefont
  {Roy}, \citenamefont {Ranga}, \citenamefont {Shoemaker}, \citenamefont
  {Song}, \citenamefont {Lundh}, \citenamefont {Choi},\ and\ \citenamefont
  {Krishnamoorthy}}]{mesfet3}%
  \BibitemOpen
  \bibfield  {author} {\bibinfo {author} {\bibfnamefont {A.}~\bibnamefont
  {Bhattacharyya}}, \bibinfo {author} {\bibfnamefont {S.}~\bibnamefont {Roy}},
  \bibinfo {author} {\bibfnamefont {P.}~\bibnamefont {Ranga}}, \bibinfo
  {author} {\bibfnamefont {D.}~\bibnamefont {Shoemaker}}, \bibinfo {author}
  {\bibfnamefont {Y.}~\bibnamefont {Song}}, \bibinfo {author} {\bibfnamefont
  {J.~S.}\ \bibnamefont {Lundh}}, \bibinfo {author} {\bibfnamefont
  {S.}~\bibnamefont {Choi}}, \ and\ \bibinfo {author} {\bibfnamefont
  {S.}~\bibnamefont {Krishnamoorthy}},\ }\bibfield  {title} {\enquote {\bibinfo
  {title} {130 ma $mm^{-1}$ $\beta-{Ga}_{2}{O}_{3}$ metal semiconductor field
  effect transistor with low-temperature metalorganic vapor phase
  epitaxy-regrown ohmic contacts},}\ }\href {\doibase
  10.35848/1882-0786/ac07ef} {\bibfield  {journal} {\bibinfo  {journal}
  {Applied Physics Express}\ }\textbf {\bibinfo {volume} {14}},\ \bibinfo
  {pages} {076502} (\bibinfo {year} {2021}{\natexlab{a}})}\BibitemShut
  {NoStop}%
\bibitem [{\citenamefont {Bhattacharyya}\ \emph
  {et~al.}(2021{\natexlab{b}})\citenamefont {Bhattacharyya}, \citenamefont
  {Ranga}, \citenamefont {Roy}, \citenamefont {Peterson}, \citenamefont
  {Alema}, \citenamefont {Seryogin}, \citenamefont {Osinsky},\ and\
  \citenamefont {Krishnamoorthy}}]{mesfet4}%
  \BibitemOpen
  \bibfield  {author} {\bibinfo {author} {\bibfnamefont {A.}~\bibnamefont
  {Bhattacharyya}}, \bibinfo {author} {\bibfnamefont {P.}~\bibnamefont
  {Ranga}}, \bibinfo {author} {\bibfnamefont {S.}~\bibnamefont {Roy}}, \bibinfo
  {author} {\bibfnamefont {C.}~\bibnamefont {Peterson}}, \bibinfo {author}
  {\bibfnamefont {F.}~\bibnamefont {Alema}}, \bibinfo {author} {\bibfnamefont
  {G.}~\bibnamefont {Seryogin}}, \bibinfo {author} {\bibfnamefont
  {A.}~\bibnamefont {Osinsky}}, \ and\ \bibinfo {author} {\bibfnamefont
  {S.}~\bibnamefont {Krishnamoorthy}},\ }\bibfield  {title} {\enquote {\bibinfo
  {title} {Multi-kv class $\beta-{Ga}_{2}{O}_{3}$ mesfets with a lateral figure
  of merit up to 355 mw/cm²},}\ }\href@noop {} {\bibfield  {journal} {\bibinfo
   {journal} {IEEE Electron Device Letters}\ }\textbf {\bibinfo {volume}
  {42}},\ \bibinfo {pages} {1272--1275} (\bibinfo {year}
  {2021}{\natexlab{b}})}\BibitemShut {NoStop}%
\bibitem [{\citenamefont {Wong}\ \emph {et~al.}(2018)\citenamefont {Wong},
  \citenamefont {Lin}, \citenamefont {Kuramata}, \citenamefont {Yamakoshi},
  \citenamefont {Murakami}, \citenamefont {Kumagai},\ and\ \citenamefont
  {Higashiwaki}}]{ptype1}%
  \BibitemOpen
  \bibfield  {author} {\bibinfo {author} {\bibfnamefont {M.~H.}\ \bibnamefont
  {Wong}}, \bibinfo {author} {\bibfnamefont {C.-H.}\ \bibnamefont {Lin}},
  \bibinfo {author} {\bibfnamefont {A.}~\bibnamefont {Kuramata}}, \bibinfo
  {author} {\bibfnamefont {S.}~\bibnamefont {Yamakoshi}}, \bibinfo {author}
  {\bibfnamefont {H.}~\bibnamefont {Murakami}}, \bibinfo {author}
  {\bibfnamefont {Y.}~\bibnamefont {Kumagai}}, \ and\ \bibinfo {author}
  {\bibfnamefont {M.}~\bibnamefont {Higashiwaki}},\ }\bibfield  {title}
  {\enquote {\bibinfo {title} {Acceptor doping of $\beta-{Ga}_{2}{O}_{3}$ by
  ${Mg}$ and ${N}$ ion implantations},}\ }\href {\doibase 10.1063/1.5050040}
  {\bibfield  {journal} {\bibinfo  {journal} {Applied Physics Letters}\
  }\textbf {\bibinfo {volume} {113}},\ \bibinfo {pages} {102103} (\bibinfo
  {year} {2018})}\BibitemShut {NoStop}%
\bibitem [{\citenamefont {Zhang}\ \emph {et~al.}(2018)\citenamefont {Zhang},
  \citenamefont {Neal}, \citenamefont {Xia}, \citenamefont {Joishi},
  \citenamefont {Johnson}, \citenamefont {Zheng}, \citenamefont {Bajaj},
  \citenamefont {Brenner}, \citenamefont {Dorsey}, \citenamefont {Chabak},
  \citenamefont {Jessen}, \citenamefont {Hwang}, \citenamefont {Mou},
  \citenamefont {Heremans},\ and\ \citenamefont {Rajan}}]{algao5}%
  \BibitemOpen
  \bibfield  {author} {\bibinfo {author} {\bibfnamefont {Y.}~\bibnamefont
  {Zhang}}, \bibinfo {author} {\bibfnamefont {A.}~\bibnamefont {Neal}},
  \bibinfo {author} {\bibfnamefont {Z.}~\bibnamefont {Xia}}, \bibinfo {author}
  {\bibfnamefont {C.}~\bibnamefont {Joishi}}, \bibinfo {author} {\bibfnamefont
  {J.~M.}\ \bibnamefont {Johnson}}, \bibinfo {author} {\bibfnamefont
  {Y.}~\bibnamefont {Zheng}}, \bibinfo {author} {\bibfnamefont
  {S.}~\bibnamefont {Bajaj}}, \bibinfo {author} {\bibfnamefont
  {M.}~\bibnamefont {Brenner}}, \bibinfo {author} {\bibfnamefont
  {D.}~\bibnamefont {Dorsey}}, \bibinfo {author} {\bibfnamefont
  {K.}~\bibnamefont {Chabak}}, \bibinfo {author} {\bibfnamefont
  {G.}~\bibnamefont {Jessen}}, \bibinfo {author} {\bibfnamefont
  {J.}~\bibnamefont {Hwang}}, \bibinfo {author} {\bibfnamefont
  {S.}~\bibnamefont {Mou}}, \bibinfo {author} {\bibfnamefont {J.~P.}\
  \bibnamefont {Heremans}}, \ and\ \bibinfo {author} {\bibfnamefont
  {S.}~\bibnamefont {Rajan}},\ }\bibfield  {title} {\enquote {\bibinfo {title}
  {Demonstration of high mobility and quantum transport in modulation-doped
  $\beta-({Al}_{x}{Ga}_{1-x})_{2}{O}_{3}/{Ga}_{2}{O}_{3}$ heterostructures},}\
  }\href {\doibase 10.1063/1.5025704} {\bibfield  {journal} {\bibinfo
  {journal} {Applied Physics Letters}\ }\textbf {\bibinfo {volume} {112}},\
  \bibinfo {pages} {173502} (\bibinfo {year} {2018})}\BibitemShut {NoStop}%
\bibitem [{\citenamefont {Kumar}, \citenamefont {Ghosh},\ and\ \citenamefont
  {Singisetti}(2020)}]{2deg1}%
  \BibitemOpen
  \bibfield  {author} {\bibinfo {author} {\bibfnamefont {A.}~\bibnamefont
  {Kumar}}, \bibinfo {author} {\bibfnamefont {K.}~\bibnamefont {Ghosh}}, \ and\
  \bibinfo {author} {\bibfnamefont {U.}~\bibnamefont {Singisetti}},\ }\bibfield
   {title} {\enquote {\bibinfo {title} {Low field transport calculation of
  2-dimensional electron gas in
  $\beta-({Al}_{x}{Ga}_{1-x})_{2}{O}_{3}/{Ga}_{2}{O}_{3}$ heterostructures},}\
  }\href {\doibase 10.1063/5.0008578} {\bibfield  {journal} {\bibinfo
  {journal} {Journal of Applied Physics}\ }\textbf {\bibinfo {volume} {128}},\
  \bibinfo {pages} {105703} (\bibinfo {year} {2020})}\BibitemShut {NoStop}%
\bibitem [{\citenamefont {Oishi}\ \emph {et~al.}(2015)\citenamefont {Oishi},
  \citenamefont {Koga}, \citenamefont {Harada},\ and\ \citenamefont
  {Kasu}}]{lfmob1}%
  \BibitemOpen
  \bibfield  {author} {\bibinfo {author} {\bibfnamefont {T.}~\bibnamefont
  {Oishi}}, \bibinfo {author} {\bibfnamefont {Y.}~\bibnamefont {Koga}},
  \bibinfo {author} {\bibfnamefont {K.}~\bibnamefont {Harada}}, \ and\ \bibinfo
  {author} {\bibfnamefont {M.}~\bibnamefont {Kasu}},\ }\bibfield  {title}
  {\enquote {\bibinfo {title} {High-mobility $\beta-{Ga}_{2}{O}_{3}$($\bar201$)
  single crystals grown by edge-defined film-fed growth method and their
  schottky barrier diodes with ni contact},}\ }\href {\doibase
  10.7567/apex.8.031101} {\bibfield  {journal} {\bibinfo  {journal} {Applied
  Physics Express}\ }\textbf {\bibinfo {volume} {8}},\ \bibinfo {pages}
  {031101} (\bibinfo {year} {2015})}\BibitemShut {NoStop}%
\bibitem [{\citenamefont {Zhang}\ \emph {et~al.}(2019)\citenamefont {Zhang},
  \citenamefont {Alema}, \citenamefont {Mauze}, \citenamefont {Koksaldi},
  \citenamefont {Miller}, \citenamefont {Osinsky},\ and\ \citenamefont
  {Speck}}]{lfmob2}%
  \BibitemOpen
  \bibfield  {author} {\bibinfo {author} {\bibfnamefont {Y.}~\bibnamefont
  {Zhang}}, \bibinfo {author} {\bibfnamefont {F.}~\bibnamefont {Alema}},
  \bibinfo {author} {\bibfnamefont {A.}~\bibnamefont {Mauze}}, \bibinfo
  {author} {\bibfnamefont {O.~S.}\ \bibnamefont {Koksaldi}}, \bibinfo {author}
  {\bibfnamefont {R.}~\bibnamefont {Miller}}, \bibinfo {author} {\bibfnamefont
  {A.}~\bibnamefont {Osinsky}}, \ and\ \bibinfo {author} {\bibfnamefont
  {J.~S.}\ \bibnamefont {Speck}},\ }\bibfield  {title} {\enquote {\bibinfo
  {title} {Mocvd grown epitaxial $\beta-{Ga}_{2}{O}_{3}$ thin film with an
  electron mobility of 176 $cm^{2}/v$ s at room temperature},}\ }\href
  {\doibase 10.1063/1.5058059} {\bibfield  {journal} {\bibinfo  {journal} {APL
  Materials}\ }\textbf {\bibinfo {volume} {7}},\ \bibinfo {pages} {022506}
  (\bibinfo {year} {2019})}\BibitemShut {NoStop}%
\bibitem [{\citenamefont {Vaidya}, \citenamefont {Saha},\ and\ \citenamefont
  {Singisetti}(2021)}]{hfet1}%
  \BibitemOpen
  \bibfield  {author} {\bibinfo {author} {\bibfnamefont {A.}~\bibnamefont
  {Vaidya}}, \bibinfo {author} {\bibfnamefont {C.~N.}\ \bibnamefont {Saha}}, \
  and\ \bibinfo {author} {\bibfnamefont {U.}~\bibnamefont {Singisetti}},\
  }\bibfield  {title} {\enquote {\bibinfo {title} {Enhancement mode
  $\beta-({Al}_{x}{Ga}_{1-x})_{2}{O}_{3}/{Ga}_{2}{O}_{3}$ {H}eterostructure
  {FET} ({HFET}) with {H}igh {T}ransconductance and {C}utoff {F}requency},}\
  }\href {\doibase 10.1109/LED.2021.3104256} {\bibfield  {journal} {\bibinfo
  {journal} {IEEE Electron Device Letters}\ }\textbf {\bibinfo {volume} {42}},\
  \bibinfo {pages} {1444--1447} (\bibinfo {year} {2021})}\BibitemShut {NoStop}%
\bibitem [{\citenamefont {Mu}, \citenamefont {Peelaers},\ and\ \citenamefont
  {Van~de Walle}(2019)}]{algao1}%
  \BibitemOpen
  \bibfield  {author} {\bibinfo {author} {\bibfnamefont {S.}~\bibnamefont
  {Mu}}, \bibinfo {author} {\bibfnamefont {H.}~\bibnamefont {Peelaers}}, \ and\
  \bibinfo {author} {\bibfnamefont {C.~G.}\ \bibnamefont {Van~de Walle}},\
  }\bibfield  {title} {\enquote {\bibinfo {title} {Ab initio study of enhanced
  thermal conductivity in ordered {AlGaO3} alloys},}\ }\href {\doibase
  10.1063/1.5131755} {\bibfield  {journal} {\bibinfo  {journal} {Applied
  Physics Letters}\ }\textbf {\bibinfo {volume} {115}},\ \bibinfo {pages}
  {242103} (\bibinfo {year} {2019})}\BibitemShut {NoStop}%
\bibitem [{\citenamefont {Peelaers}\ \emph {et~al.}(2018)\citenamefont
  {Peelaers}, \citenamefont {Varley}, \citenamefont {Speck},\ and\
  \citenamefont {Van~de Walle}}]{algao2}%
  \BibitemOpen
  \bibfield  {author} {\bibinfo {author} {\bibfnamefont {H.}~\bibnamefont
  {Peelaers}}, \bibinfo {author} {\bibfnamefont {J.~B.}\ \bibnamefont
  {Varley}}, \bibinfo {author} {\bibfnamefont {J.~S.}\ \bibnamefont {Speck}}, \
  and\ \bibinfo {author} {\bibfnamefont {C.~G.}\ \bibnamefont {Van~de Walle}},\
  }\bibfield  {title} {\enquote {\bibinfo {title} {Structural and electronic
  properties of ${Ga}_{2}{O}_{3}-{Al}_{2}{O}_{3}$ alloys},}\ }\href {\doibase
  10.1063/1.5036991} {\bibfield  {journal} {\bibinfo  {journal} {Applied
  Physics Letters}\ }\textbf {\bibinfo {volume} {112}},\ \bibinfo {pages}
  {242101} (\bibinfo {year} {2018})}\BibitemShut {NoStop}%
\bibitem [{\citenamefont {Varley}\ \emph {et~al.}(2020)\citenamefont {Varley},
  \citenamefont {Perron}, \citenamefont {Lordi}, \citenamefont
  {Wickramaratne},\ and\ \citenamefont {Lyons}}]{algao3}%
  \BibitemOpen
  \bibfield  {author} {\bibinfo {author} {\bibfnamefont {J.~B.}\ \bibnamefont
  {Varley}}, \bibinfo {author} {\bibfnamefont {A.}~\bibnamefont {Perron}},
  \bibinfo {author} {\bibfnamefont {V.}~\bibnamefont {Lordi}}, \bibinfo
  {author} {\bibfnamefont {D.}~\bibnamefont {Wickramaratne}}, \ and\ \bibinfo
  {author} {\bibfnamefont {J.~L.}\ \bibnamefont {Lyons}},\ }\bibfield  {title}
  {\enquote {\bibinfo {title} {Prospects for n-type doping of (alxga1-x)2o3
  alloys},}\ }\href {\doibase 10.1063/5.0006224} {\bibfield  {journal}
  {\bibinfo  {journal} {Applied Physics Letters}\ }\textbf {\bibinfo {volume}
  {116}},\ \bibinfo {pages} {172104} (\bibinfo {year} {2020})}\BibitemShut
  {NoStop}%
\bibitem [{\citenamefont {Anhar Uddin~Bhuiyan}\ \emph
  {et~al.}(2019)\citenamefont {Anhar Uddin~Bhuiyan}, \citenamefont {Feng},
  \citenamefont {Johnson}, \citenamefont {Chen}, \citenamefont {Huang},
  \citenamefont {Hwang},\ and\ \citenamefont {Zhao}}]{algao4}%
  \BibitemOpen
  \bibfield  {author} {\bibinfo {author} {\bibfnamefont {A.~F.~M.}\
  \bibnamefont {Anhar Uddin~Bhuiyan}}, \bibinfo {author} {\bibfnamefont
  {Z.}~\bibnamefont {Feng}}, \bibinfo {author} {\bibfnamefont {J.~M.}\
  \bibnamefont {Johnson}}, \bibinfo {author} {\bibfnamefont {Z.}~\bibnamefont
  {Chen}}, \bibinfo {author} {\bibfnamefont {H.-L.}\ \bibnamefont {Huang}},
  \bibinfo {author} {\bibfnamefont {J.}~\bibnamefont {Hwang}}, \ and\ \bibinfo
  {author} {\bibfnamefont {H.}~\bibnamefont {Zhao}},\ }\bibfield  {title}
  {\enquote {\bibinfo {title} {{MOCVD} epitaxy of beta-(alxga1-x)2o3 thin films
  on (010) ga2o3 substrates and n-type doping},}\ }\href {\doibase
  10.1063/1.5123495} {\bibfield  {journal} {\bibinfo  {journal} {Applied
  Physics Letters}\ }\textbf {\bibinfo {volume} {115}},\ \bibinfo {pages}
  {120602} (\bibinfo {year} {2019})}\BibitemShut {NoStop}%
\bibitem [{\citenamefont {Nordheim}(1931)}]{vca}%
  \BibitemOpen
  \bibfield  {author} {\bibinfo {author} {\bibfnamefont {L.}~\bibnamefont
  {Nordheim}},\ }\bibfield  {title} {\enquote {\bibinfo {title} {Zur
  elektronentheorie der metalle. i},}\ }\href {\doibase
  https://doi.org/10.1002/andp.19314010507} {\bibfield  {journal} {\bibinfo
  {journal} {Annalen der Physik}\ }\textbf {\bibinfo {volume} {401}},\ \bibinfo
  {pages} {607--640} (\bibinfo {year} {1931})}\BibitemShut {NoStop}%
\bibitem [{\citenamefont {Yonezawa}\ and\ \citenamefont
  {Morigaki}(1973)}]{cpa}%
  \BibitemOpen
  \bibfield  {author} {\bibinfo {author} {\bibfnamefont {F.}~\bibnamefont
  {Yonezawa}}\ and\ \bibinfo {author} {\bibfnamefont {K.}~\bibnamefont
  {Morigaki}},\ }\bibfield  {title} {\enquote {\bibinfo {title} {{Coherent
  {P}otential {A}pproximation. {B}asic concepts and applications}},}\ }\href
  {\doibase 10.1143/PTPS.53.1} {\bibfield  {journal} {\bibinfo  {journal}
  {Progress of Theoretical Physics Supplement}\ }\textbf {\bibinfo {volume}
  {53}},\ \bibinfo {pages} {1--76} (\bibinfo {year} {1973})}\BibitemShut
  {NoStop}%
\bibitem [{\citenamefont {Yukinobu}(2000)}]{cpa1}%
  \BibitemOpen
  \bibfield  {author} {\bibinfo {author} {\bibfnamefont {K.}~\bibnamefont
  {Yukinobu}},\ }\href@noop {} {\emph {\bibinfo {title} {Electric Refractory
  Material}}}\ (\bibinfo  {publisher} {CRC Press},\ \bibinfo {year}
  {2000})\BibitemShut {NoStop}%
\bibitem [{\citenamefont {Meike}\ \emph {et~al.}(2000)\citenamefont {Meike},
  \citenamefont {Gonis}, \citenamefont {Turchi},\ and\ \citenamefont
  {Rajan}}]{cpa2}%
  \BibitemOpen
  \bibfield  {author} {\bibinfo {author} {\bibfnamefont {A.}~\bibnamefont
  {Meike}}, \bibinfo {author} {\bibfnamefont {A.}~\bibnamefont {Gonis}},
  \bibinfo {author} {\bibfnamefont {P.~E.}\ \bibnamefont {Turchi}}, \ and\
  \bibinfo {author} {\bibfnamefont {K.}~\bibnamefont {Rajan}},\ }\href@noop {}
  {\emph {\bibinfo {title} {Properties of Complex Inorganic Solids}}}\
  (\bibinfo  {publisher} {Springer},\ \bibinfo {year} {2000})\BibitemShut
  {NoStop}%
\bibitem [{\citenamefont {Popescu}\ and\ \citenamefont
  {Zunger}(2012)}]{bandunfold1}%
  \BibitemOpen
  \bibfield  {author} {\bibinfo {author} {\bibfnamefont {V.}~\bibnamefont
  {Popescu}}\ and\ \bibinfo {author} {\bibfnamefont {A.}~\bibnamefont
  {Zunger}},\ }\bibfield  {title} {\enquote {\bibinfo {title} {Extracting
  $\textbf{E}$ versus $\vec{\textbf{k}}$ effective band structure from
  supercell calculations on alloys and impurities},}\ }\href {\doibase
  10.1103/PhysRevB.85.085201} {\bibfield  {journal} {\bibinfo  {journal} {Phys.
  Rev. B}\ }\textbf {\bibinfo {volume} {85}},\ \bibinfo {pages} {085201}
  (\bibinfo {year} {2012})}\BibitemShut {NoStop}%
\bibitem [{\citenamefont {Popescu}\ and\ \citenamefont
  {Zunger}(2010)}]{bandunfold2}%
  \BibitemOpen
  \bibfield  {author} {\bibinfo {author} {\bibfnamefont {V.}~\bibnamefont
  {Popescu}}\ and\ \bibinfo {author} {\bibfnamefont {A.}~\bibnamefont
  {Zunger}},\ }\bibfield  {title} {\enquote {\bibinfo {title} {Effective band
  structure of random alloys},}\ }\href {\doibase
  10.1103/PhysRevLett.104.236403} {\bibfield  {journal} {\bibinfo  {journal}
  {Phys. Rev. Lett.}\ }\textbf {\bibinfo {volume} {104}},\ \bibinfo {pages}
  {236403} (\bibinfo {year} {2010})}\BibitemShut {NoStop}%
\bibitem [{\citenamefont {Wang}\ \emph {et~al.}(2018)\citenamefont {Wang},
  \citenamefont {Li}, \citenamefont {Ni},\ and\ \citenamefont
  {Janotti}}]{algaovol1}%
  \BibitemOpen
  \bibfield  {author} {\bibinfo {author} {\bibfnamefont {T.}~\bibnamefont
  {Wang}}, \bibinfo {author} {\bibfnamefont {W.}~\bibnamefont {Li}}, \bibinfo
  {author} {\bibfnamefont {C.}~\bibnamefont {Ni}}, \ and\ \bibinfo {author}
  {\bibfnamefont {A.}~\bibnamefont {Janotti}},\ }\bibfield  {title} {\enquote
  {\bibinfo {title} {Band gap and band offset of ${Ga}_{2}{O}_{3}$ and
  $({Al}_{x}{Ga}_{1-x})_{2}{O}_{3}$ alloys},}\ }\href {\doibase
  10.1103/PhysRevApplied.10.011003} {\bibfield  {journal} {\bibinfo  {journal}
  {Phys. Rev. Applied}\ }\textbf {\bibinfo {volume} {10}},\ \bibinfo {pages}
  {011003} (\bibinfo {year} {2018})}\BibitemShut {NoStop}%
\bibitem [{\citenamefont {Zunger}\ \emph {et~al.}(1990)\citenamefont {Zunger},
  \citenamefont {Wei}, \citenamefont {Ferreira},\ and\ \citenamefont
  {Bernard}}]{sqs1}%
  \BibitemOpen
  \bibfield  {author} {\bibinfo {author} {\bibfnamefont {A.}~\bibnamefont
  {Zunger}}, \bibinfo {author} {\bibfnamefont {S.-H.}\ \bibnamefont {Wei}},
  \bibinfo {author} {\bibfnamefont {L.~G.}\ \bibnamefont {Ferreira}}, \ and\
  \bibinfo {author} {\bibfnamefont {J.~E.}\ \bibnamefont {Bernard}},\
  }\bibfield  {title} {\enquote {\bibinfo {title} {Special quasirandom
  structures},}\ }\href {\doibase 10.1103/PhysRevLett.65.353} {\bibfield
  {journal} {\bibinfo  {journal} {Phys. Rev. Lett.}\ }\textbf {\bibinfo
  {volume} {65}},\ \bibinfo {pages} {353--356} (\bibinfo {year}
  {1990})}\BibitemShut {NoStop}%
\bibitem [{\citenamefont {Wei}\ \emph {et~al.}(1990)\citenamefont {Wei},
  \citenamefont {Ferreira}, \citenamefont {Bernard},\ and\ \citenamefont
  {Zunger}}]{sqs2}%
  \BibitemOpen
  \bibfield  {author} {\bibinfo {author} {\bibfnamefont {S.-H.}\ \bibnamefont
  {Wei}}, \bibinfo {author} {\bibfnamefont {L.~G.}\ \bibnamefont {Ferreira}},
  \bibinfo {author} {\bibfnamefont {J.~E.}\ \bibnamefont {Bernard}}, \ and\
  \bibinfo {author} {\bibfnamefont {A.}~\bibnamefont {Zunger}},\ }\bibfield
  {title} {\enquote {\bibinfo {title} {Electronic properties of random alloys:
  Special quasirandom structures},}\ }\href {\doibase 10.1103/PhysRevB.42.9622}
  {\bibfield  {journal} {\bibinfo  {journal} {Phys. Rev. B}\ }\textbf {\bibinfo
  {volume} {42}},\ \bibinfo {pages} {9622--9649} (\bibinfo {year}
  {1990})}\BibitemShut {NoStop}%
\bibitem [{\citenamefont {van~de Walle}\ \emph {et~al.}(2013)\citenamefont
  {van~de Walle}, \citenamefont {Tiwary}, \citenamefont {de~Jong},
  \citenamefont {Olmsted}, \citenamefont {Asta}, \citenamefont {Dick},
  \citenamefont {Shin}, \citenamefont {Wang}, \citenamefont {Chen},\ and\
  \citenamefont {Liu}}]{atat2}%
  \BibitemOpen
  \bibfield  {author} {\bibinfo {author} {\bibfnamefont {A.}~\bibnamefont
  {van~de Walle}}, \bibinfo {author} {\bibfnamefont {P.}~\bibnamefont
  {Tiwary}}, \bibinfo {author} {\bibfnamefont {M.~M.}\ \bibnamefont {de~Jong}},
  \bibinfo {author} {\bibfnamefont {D.~L.}\ \bibnamefont {Olmsted}}, \bibinfo
  {author} {\bibfnamefont {M.~D.}\ \bibnamefont {Asta}}, \bibinfo {author}
  {\bibfnamefont {A.}~\bibnamefont {Dick}}, \bibinfo {author} {\bibfnamefont
  {D.}~\bibnamefont {Shin}}, \bibinfo {author} {\bibfnamefont {Y.}~\bibnamefont
  {Wang}}, \bibinfo {author} {\bibfnamefont {L.-Q.}\ \bibnamefont {Chen}}, \
  and\ \bibinfo {author} {\bibfnamefont {Z.-K.}\ \bibnamefont {Liu}},\
  }\bibfield  {title} {\enquote {\bibinfo {title} {Efficient stochastic
  generation of special quasirandom structures},}\ }\href {\doibase
  10.1016/j.calphad.2013.06.006} {\bibfield  {journal} {\bibinfo  {journal}
  {Calphad}\ }\textbf {\bibinfo {volume} {42}},\ \bibinfo {pages} {13--18}
  (\bibinfo {year} {2013})}\BibitemShut {NoStop}%
\bibitem [{\citenamefont {van~de Walle}(2009)}]{atat3}%
  \BibitemOpen
  \bibfield  {author} {\bibinfo {author} {\bibfnamefont {A.}~\bibnamefont
  {van~de Walle}},\ }\bibfield  {title} {\enquote {\bibinfo {title}
  {{M}ulticomponent multisublattice alloys, nonconfigurational entropy and
  other additions to the {A}lloy {T}heoretic {A}utomated {T}oolkit},}\ }\href
  {\doibase 10.1016/j.calphad.2008.12.005} {\bibfield  {journal} {\bibinfo
  {journal} {Calphad}\ }\textbf {\bibinfo {volume} {33}},\ \bibinfo {pages}
  {266--278} (\bibinfo {year} {2009})}\BibitemShut {NoStop}%
\bibitem [{\citenamefont {Kresse}\ and\ \citenamefont {Hafner}(1993)}]{vasp1}%
  \BibitemOpen
  \bibfield  {author} {\bibinfo {author} {\bibfnamefont {G.}~\bibnamefont
  {Kresse}}\ and\ \bibinfo {author} {\bibfnamefont {J.}~\bibnamefont
  {Hafner}},\ }\bibfield  {title} {\enquote {\bibinfo {title} {Ab initio
  molecular dynamics for liquid metals},}\ }\href {\doibase
  10.1103/PhysRevB.47.558} {\bibfield  {journal} {\bibinfo  {journal} {Phys.
  Rev. B}\ }\textbf {\bibinfo {volume} {47}},\ \bibinfo {pages} {558--561}
  (\bibinfo {year} {1993})}\BibitemShut {NoStop}%
\bibitem [{\citenamefont {Kresse}\ and\ \citenamefont
  {Furthmüller}(1996)}]{vasp2}%
  \BibitemOpen
  \bibfield  {author} {\bibinfo {author} {\bibfnamefont {G.}~\bibnamefont
  {Kresse}}\ and\ \bibinfo {author} {\bibfnamefont {J.}~\bibnamefont
  {Furthmüller}},\ }\bibfield  {title} {\enquote {\bibinfo {title} {Efficiency
  of ab-initio total energy calculations for metals and semiconductors using a
  plane-wave basis set},}\ }\href {\doibase
  https://doi.org/10.1016/0927-0256(96)00008-0} {\bibfield  {journal} {\bibinfo
   {journal} {Computational Materials Science}\ }\textbf {\bibinfo {volume}
  {6}},\ \bibinfo {pages} {15--50} (\bibinfo {year} {1996})}\BibitemShut
  {NoStop}%
\bibitem [{\citenamefont {Kresse}\ and\ \citenamefont
  {Furthm\"uller}(1996)}]{vasp3}%
  \BibitemOpen
  \bibfield  {author} {\bibinfo {author} {\bibfnamefont {G.}~\bibnamefont
  {Kresse}}\ and\ \bibinfo {author} {\bibfnamefont {J.}~\bibnamefont
  {Furthm\"uller}},\ }\bibfield  {title} {\enquote {\bibinfo {title} {Efficient
  iterative schemes for ab initio total-energy calculations using a plane-wave
  basis set},}\ }\href {\doibase 10.1103/PhysRevB.54.11169} {\bibfield
  {journal} {\bibinfo  {journal} {Phys. Rev. B}\ }\textbf {\bibinfo {volume}
  {54}},\ \bibinfo {pages} {11169--11186} (\bibinfo {year} {1996})}\BibitemShut
  {NoStop}%
\bibitem [{\citenamefont {Kresse}\ and\ \citenamefont {Joubert}(1999)}]{paw1}%
  \BibitemOpen
  \bibfield  {author} {\bibinfo {author} {\bibfnamefont {G.}~\bibnamefont
  {Kresse}}\ and\ \bibinfo {author} {\bibfnamefont {D.}~\bibnamefont
  {Joubert}},\ }\bibfield  {title} {\enquote {\bibinfo {title} {From ultrasoft
  pseudopotentials to the projector augmented-wave method},}\ }\href {\doibase
  10.1103/PhysRevB.59.1758} {\bibfield  {journal} {\bibinfo  {journal} {Phys.
  Rev. B}\ }\textbf {\bibinfo {volume} {59}},\ \bibinfo {pages} {1758--1775}
  (\bibinfo {year} {1999})}\BibitemShut {NoStop}%
\bibitem [{\citenamefont {Peelaers}\ and\ \citenamefont {Van~de
  Walle}(2015)}]{bz}%
  \BibitemOpen
  \bibfield  {author} {\bibinfo {author} {\bibfnamefont {H.}~\bibnamefont
  {Peelaers}}\ and\ \bibinfo {author} {\bibfnamefont {C.~G.}\ \bibnamefont
  {Van~de Walle}},\ }\bibfield  {title} {\enquote {\bibinfo {title} {Brillouin
  zone and band structure of $\beta-{Ga}_{2}{O}_{3}$},}\ }\href {\doibase
  https://doi.org/10.1002/pssb.201451551} {\bibfield  {journal} {\bibinfo
  {journal} {physica status solidi (b)}\ }\textbf {\bibinfo {volume} {252}},\
  \bibinfo {pages} {828--832} (\bibinfo {year} {2015})}\BibitemShut {NoStop}%
\bibitem [{\citenamefont {Pant}, \citenamefont {Deng},\ and\ \citenamefont
  {Kioupakis}(2020)}]{alloyscat1}%
  \BibitemOpen
  \bibfield  {author} {\bibinfo {author} {\bibfnamefont {N.}~\bibnamefont
  {Pant}}, \bibinfo {author} {\bibfnamefont {Z.}~\bibnamefont {Deng}}, \ and\
  \bibinfo {author} {\bibfnamefont {E.}~\bibnamefont {Kioupakis}},\ }\bibfield
  {title} {\enquote {\bibinfo {title} {High electron mobility of alxga1-xn
  evaluated by unfolding the dft band structure},}\ }\href {\doibase
  10.1063/5.0027802} {\bibfield  {journal} {\bibinfo  {journal} {Applied
  Physics Letters}\ }\textbf {\bibinfo {volume} {117}},\ \bibinfo {pages}
  {242105} (\bibinfo {year} {2020})}\BibitemShut {NoStop}%
\bibitem [{\citenamefont {Ghosh}\ and\ \citenamefont
  {Singisetti}(2016)}]{aniso1}%
  \BibitemOpen
  \bibfield  {author} {\bibinfo {author} {\bibfnamefont {K.}~\bibnamefont
  {Ghosh}}\ and\ \bibinfo {author} {\bibfnamefont {U.}~\bibnamefont
  {Singisetti}},\ }\bibfield  {title} {\enquote {\bibinfo {title} {Ab initio
  calculation of electron–phonon coupling in monoclinic
  $\beta-{Ga}_{2}{O}_{3}$ crystal},}\ }\href {\doibase 10.1063/1.4961308}
  {\bibfield  {journal} {\bibinfo  {journal} {Applied Physics Letters}\
  }\textbf {\bibinfo {volume} {109}},\ \bibinfo {pages} {072102} (\bibinfo
  {year} {2016})}\BibitemShut {NoStop}%
\bibitem [{\citenamefont {Perdew}, \citenamefont {Burke},\ and\ \citenamefont
  {Ernzerhof}(1996)}]{pbe}%
  \BibitemOpen
  \bibfield  {author} {\bibinfo {author} {\bibfnamefont {J.~P.}\ \bibnamefont
  {Perdew}}, \bibinfo {author} {\bibfnamefont {K.}~\bibnamefont {Burke}}, \
  and\ \bibinfo {author} {\bibfnamefont {M.}~\bibnamefont {Ernzerhof}},\
  }\bibfield  {title} {\enquote {\bibinfo {title} {Generalized {G}radient
  {A}pproximation {M}ade {S}imple},}\ }\href {\doibase
  10.1103/PhysRevLett.77.3865} {\bibfield  {journal} {\bibinfo  {journal}
  {Phys. Rev. Lett.}\ }\textbf {\bibinfo {volume} {77}},\ \bibinfo {pages}
  {3865--3868} (\bibinfo {year} {1996})}\BibitemShut {NoStop}%
\bibitem [{\citenamefont {Momma}\ and\ \citenamefont {Izumi}(2011)}]{vesta}%
  \BibitemOpen
  \bibfield  {author} {\bibinfo {author} {\bibfnamefont {K.}~\bibnamefont
  {Momma}}\ and\ \bibinfo {author} {\bibfnamefont {F.}~\bibnamefont {Izumi}},\
  }\bibfield  {title} {\enquote {\bibinfo {title} {{{\it VESTA3} for
  three-dimensional visualization of crystal, volumetric and morphology
  data}},}\ }\href {\doibase 10.1107/S0021889811038970} {\bibfield  {journal}
  {\bibinfo  {journal} {Journal of Applied Crystallography}\ }\textbf {\bibinfo
  {volume} {44}},\ \bibinfo {pages} {1272--1276} (\bibinfo {year}
  {2011})}\BibitemShut {NoStop}%
\end{thebibliography}%

\newpage

\begin{figure*}
    \centering
    \includegraphics{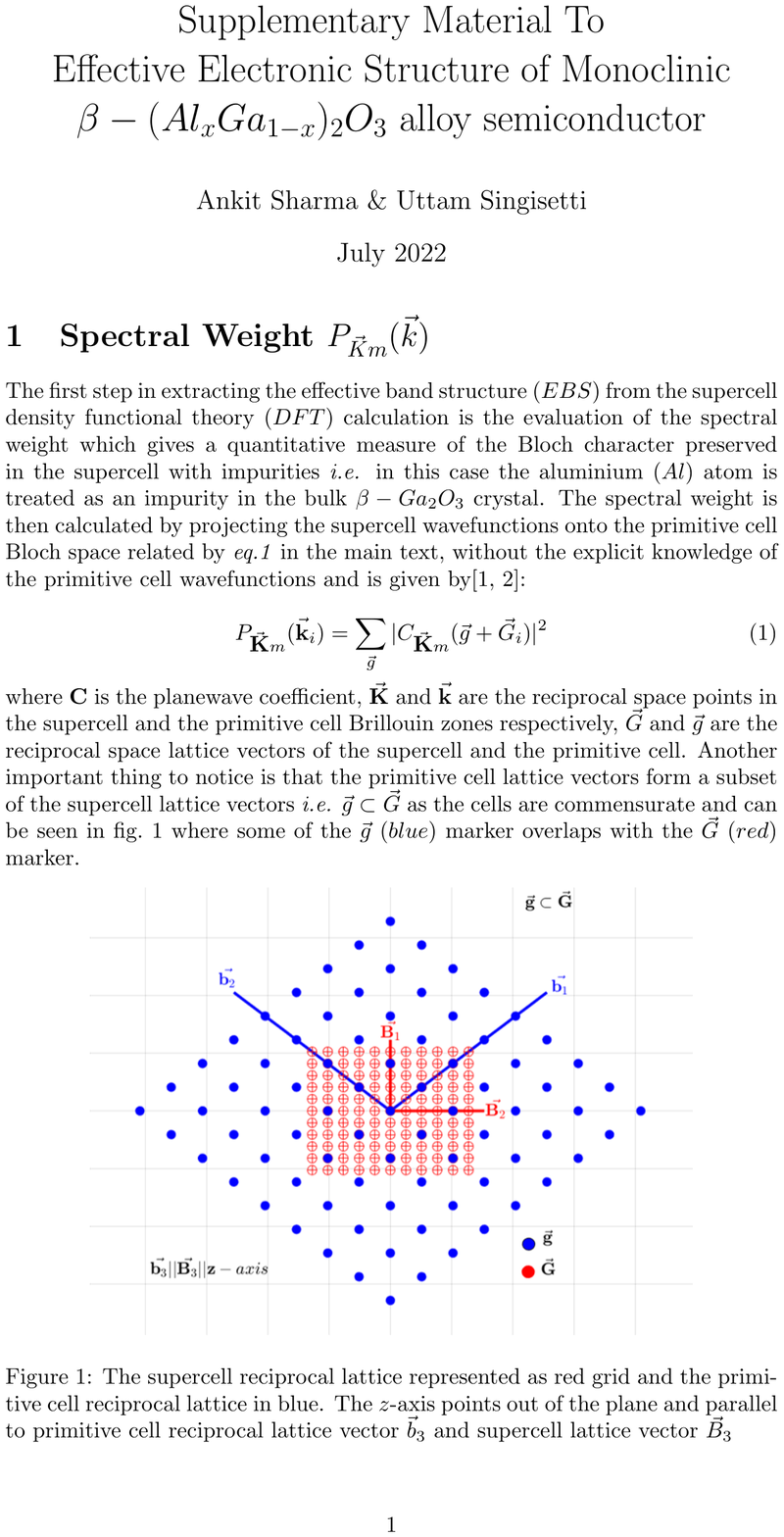}
\end{figure*}

\begin{figure*}
    \centering
    \includegraphics{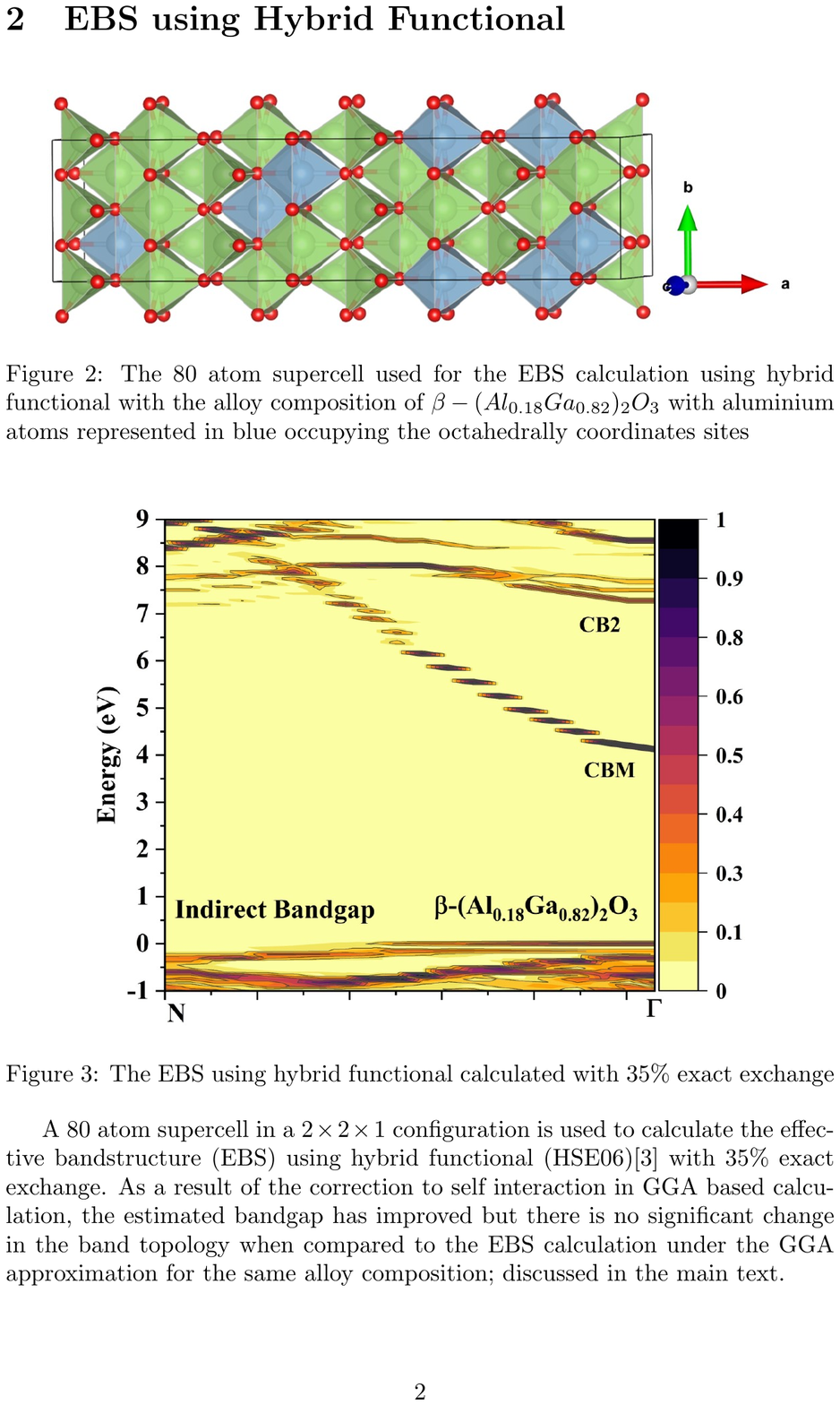}
\end{figure*}

\begin{figure*}
    \centering
    \includegraphics{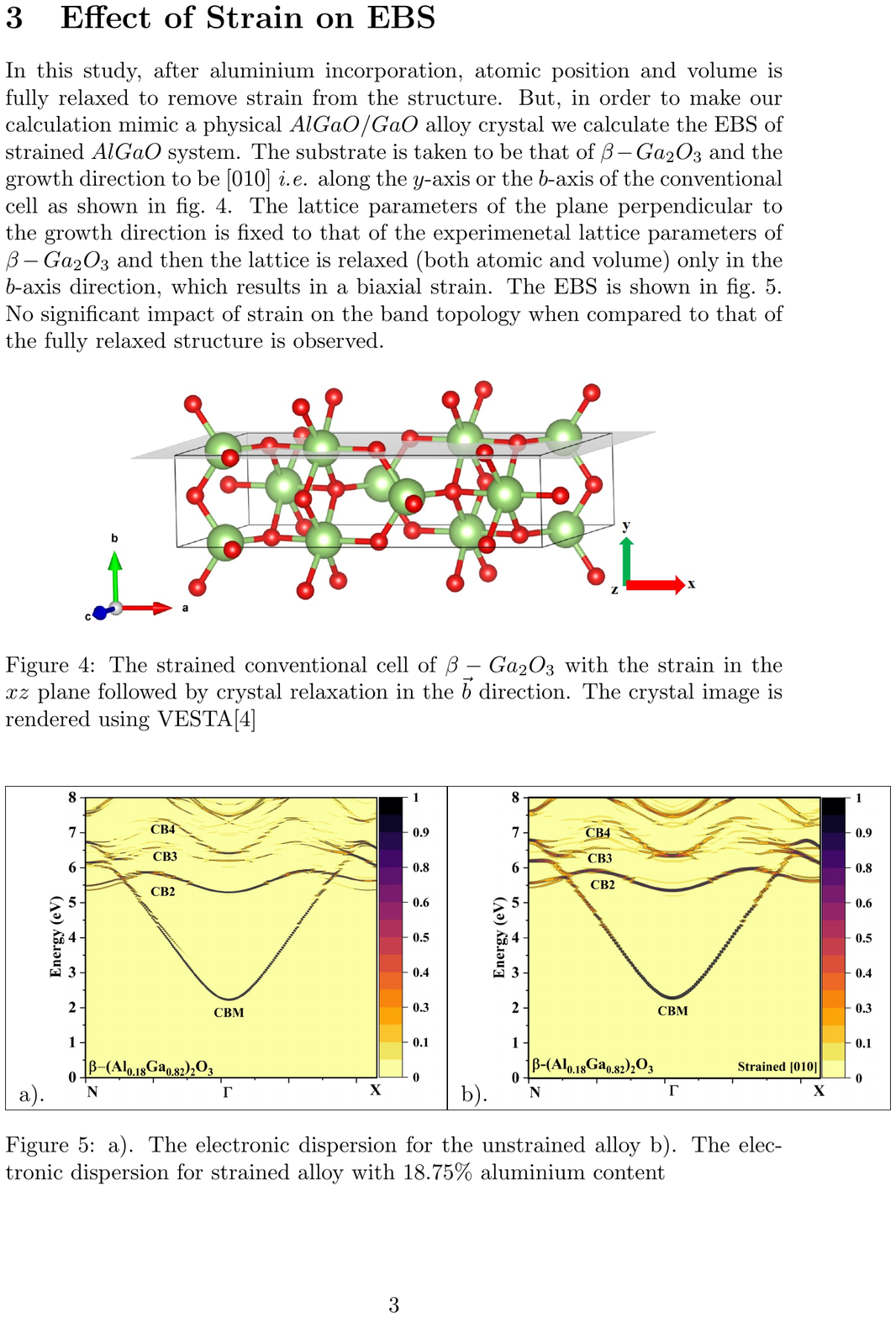}
\end{figure*}

\begin{figure*}
    \centering
    \includegraphics{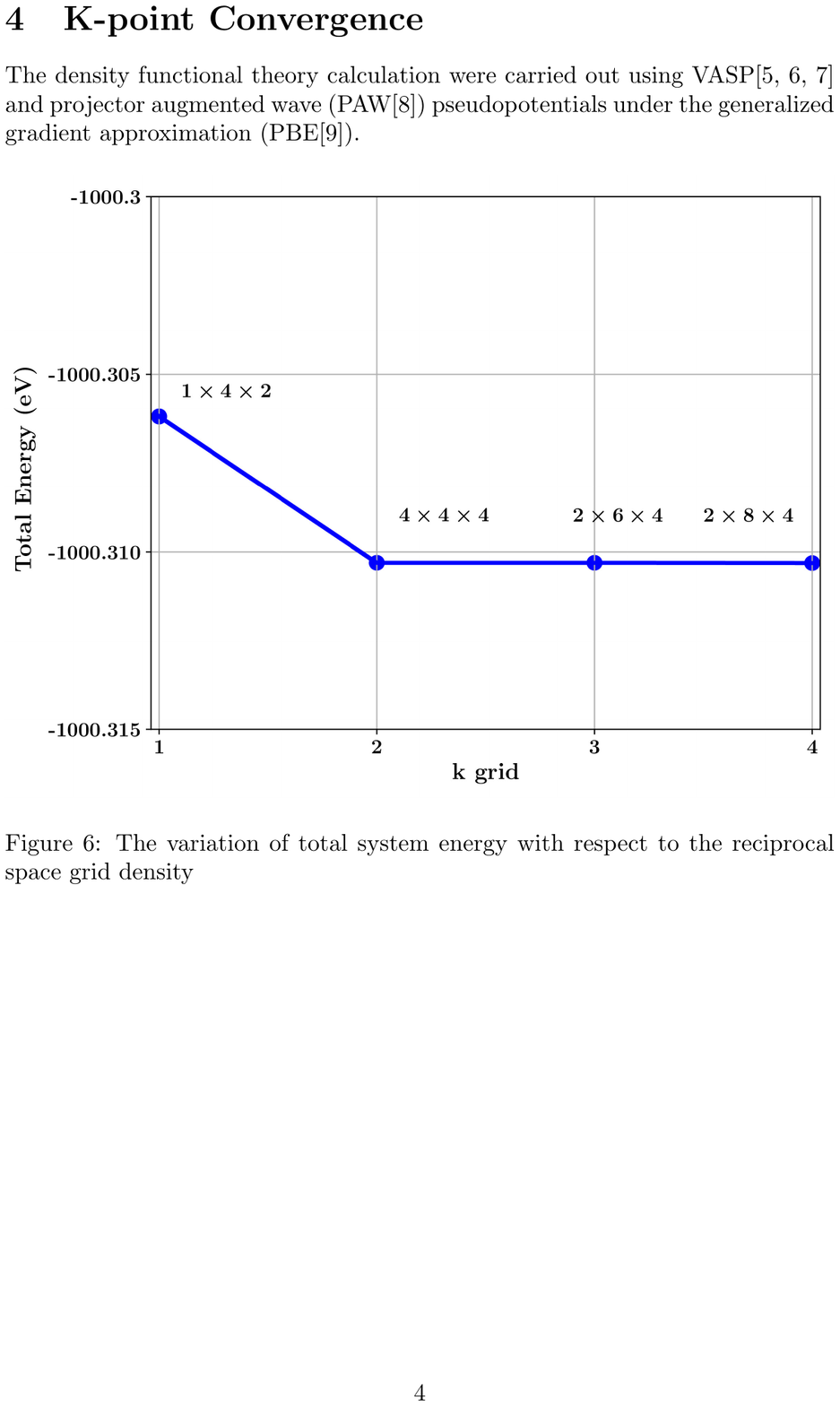}
\end{figure*}

\begin{figure*}
    \centering
    \includegraphics{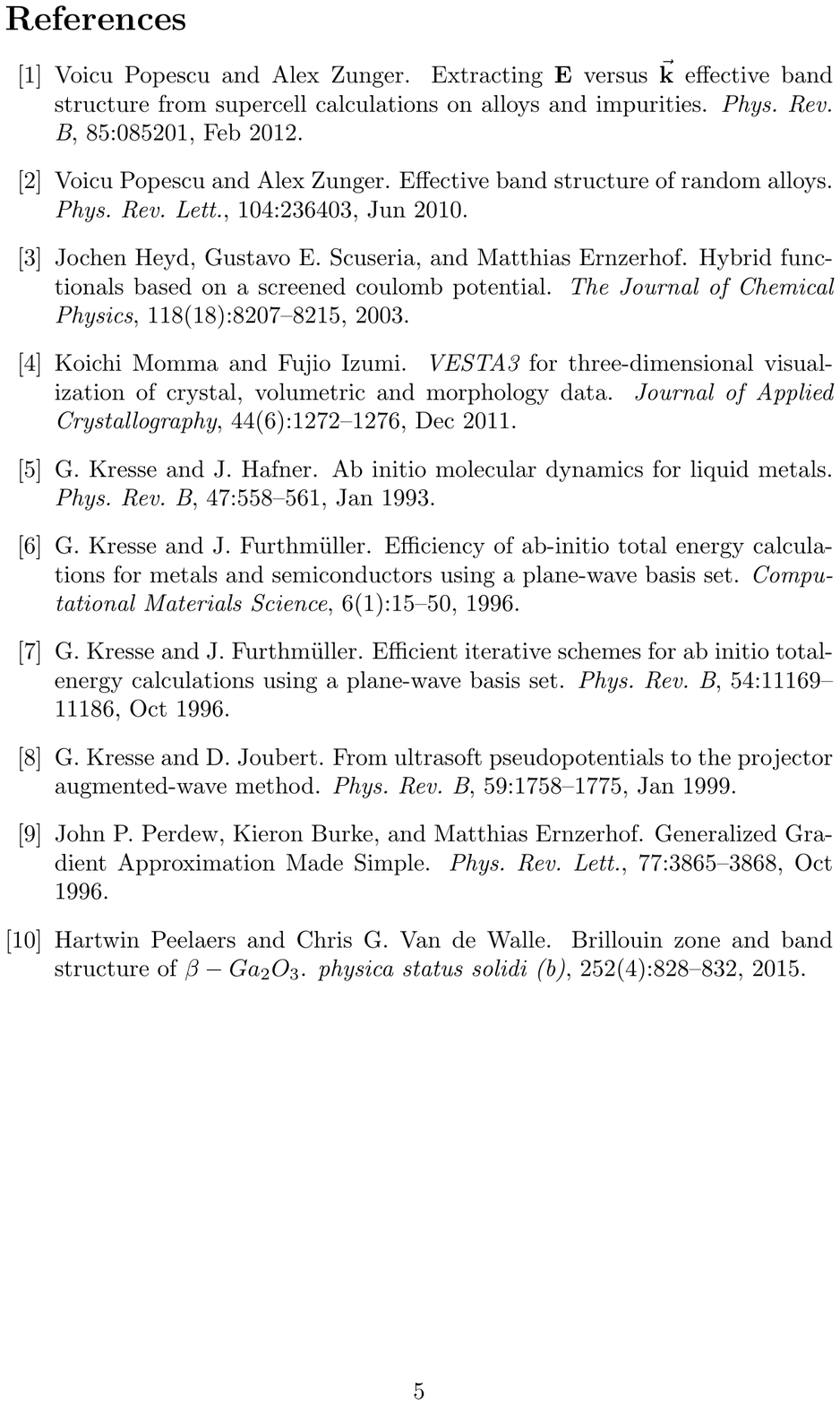}
\end{figure*}

\end{document}